\documentclass[5p,twocolumn,sort&compress,authoryear,]{elsarticle}

\usepackage{amssymb}
\usepackage{amsthm}
\usepackage{amsmath}
\usepackage{amsfonts}
\usepackage{algorithmic}
\usepackage{tcolorbox}
\usepackage{colortbl}
\usepackage{tabularx}
\usepackage{multirow}
\usepackage{stmaryrd}
\usepackage{lscape}
\usepackage{setspace}
\usepackage{subcaption}
\usepackage{placeins}
\usepackage{changepage}
\usepackage{enumitem}
\usepackage{longtable,url}
\usepackage{soul}

\theoremstyle{definition}
\newtheorem{definition}{Definition}
\theoremstyle{plain}

\theoremstyle{remark}

\newtheorem*{example*}{Example}

\usepackage{lineno}
\journal{Journal of Systems and Software}

\begin{document}

\begin{frontmatter}



\title{A Compositional Approach to Creating Architecture Frameworks \\ with an Application to Distributed AI Systems}


\author[inst1]{Hans-Martin Heyn}
\author[inst1]{Eric Knauss}
\affiliation[inst1]{organization={Department of Computer Science and Engineering, University of Gothenburg and Chalmers},
            city={Gothenburg},
            postcode={41756}, 
            country={Sweden}}

\author[inst2]{Patrizio Pelliccione}
\affiliation[inst2]{organization={Gran Sasso Science Institute (GSSI)},
            city={L'Aquila},
            postcode={67100}, 
            country={Italy}}

\begin{abstract}
Artificial intelligence (AI) in its various forms finds more and more its way into complex distributed systems. For instance, it is used locally, as part of a sensor system, on the edge for low-latency high-performance inference, or in the cloud, e.g. for data mining. 
Modern complex systems, such as connected vehicles, are often part of an Internet of Things (IoT). This poses additional architectural challenges. To manage complexity, architectures are described with architecture frameworks, which are composed of a number of architectural views connected through correspondence rules. 
Despite some attempts, the definition of a mathematical foundation for architecture frameworks that are suitable for the development of distributed AI systems still requires investigation and study.

\noindent In this paper, we propose to extend the state of the art on architecture framework by providing a mathematical model for system architectures, which is scalable and supports co-evolution of different aspects {for example} of an AI system.
Based on Design Science Research, this study starts by identifying the challenges with architectural frameworks in a use case of distributed AI systems. Then, we derive from the identified challenges four rules, and we formulate them by exploiting concepts from category theory. We show how compositional thinking can provide rules for the creation and management of architectural frameworks for complex systems, for example distributed systems with AI. The aim of the paper is not to provide viewpoints or architecture models specific to AI systems, but instead to provide guidelines based on a mathematical formulation on how a consistent framework can be built up with existing, or newly created, viewpoints.
To put in practice and test the approach, the identified and formulated rules are applied to derive an architectural framework for the EU Horizon 2020 project ``Very efficient deep learning in the IoT" (VEDLIoT) in the form of a case study. 

\end{abstract}





\end{frontmatter}


\section{Introduction}
Architectural frameworks provide knowledge structures that allow for the division of architectural descriptions into different architectural views~\citep{Pelliccione2017}. 
An architectural view expresses ``the architecture of a system from the perspective of specific system concern"~\citep{ISO42010}. 
The conventions of how an architectural view is constructed and interpreted is given through a corresponding architectural viewpoint. 
The design of a system-of-interest needs to account for different concerns of different stakeholders. 
Therefore, the architecture of the system-of-interest must be expressed through many different architectural views.\par

Designing a large and distributed system is a hierarchical process~\citep{Murugesan2019}.
Several views of the architecture of the system-of-interest allow for decomposing the design task into smaller and specialised tasks.
This hierarchical design process allows for the co-evolution of requirements and architecture, known as the ``twin peaks of requirements and architecture", as described in~\citet{Nuseibeh2001,Cleland-Huang2013}. 

Developing a complex system, which can include some form of AI, is a highly collaborative act (known as co-design) between many stakeholders. 
However, the term \emph{co-design} can have two meanings. 
It can be an acronym for \emph{collaborative design}, which means that all required stakeholders, i.e., developers of different disciplines, customers, business owners, etc, are being heard, and in some form actively involved in system design process \citet{Fitzgerald2014,Nalchigar2021}. 
Co-design can also stand for \emph{integrated design} of a system, which means that design aspects of the system, e.g., hardware, software but also quality aspects are closely coupled to each other.
In fact, many different concerns need to be satisfied with a system that includes a huge variety of different dimensions which need to be designed in parallel to ensure a ``safe by design", ``secure by design", ``ethical by design", or any other required ``quality by design" system.\par

Architecture frameworks can explicitly support various non-functional quality requirements. 
Examples include a framework for addressing non-functional requirements in computer vision systems with AI \citep{Fenn2016}, a framework for architectural patterns improving operational stability (robustness) of machine learning systems \citep{Yokoyama2019}, a framework for architectures of machine learning enabled systems that improve safety \citep{Serban2019}, and a framework for architecture frameworks of distributed AI systems that emphasise security aspects \citep{Mendhurwar2021}.\par
Each of these examples presents an architecture for AI systems which is tailored to only one particular quality aspect. In fact, architectural frameworks for AI systems are typically defined for one specific application domain. Examples of domain specific architectural framework include a framework for computer vision and cognition \citep{Kurup2011}, for self-driving vehicles \citep{Schroeder2015}, for smart power grid \citep{Thilakarathne2020}, and for health systems applications with machine learning \citep{Moreb2020}. In a systematic literature review on software engineering patterns for AI, \citet{Martinez2021} noted that ``at the system level, there are few proposals for patterns, design standards, or reference architectures". 
Also, in a review on architecture and design patterns for designing machine learning systems, \citet{Washizaki2019} noted that ``developers are concerned with the complexity of ML systems and their lack of knowledge of the architecture[...]".\par

Summarising, we believe that the following three limitations of current architectural frameworks, especially towards AI systems, exist: 

\begin{itemize}
    \item For complex systems, such as for example AI systems distributed in the IoT, there is the need of both interpretations of co-design, i.e., collaborative design and integrated design to arrive at the desired system.
    \item If quality aspects are considered in a system architecture, mostly only certain specific non-functional requirements are adopted (such as security, or safety, or explainability). Again, a more generalisable approach is needed that can include any number of required quality aspects of the system.
    \item Many architecture frameworks are limited to specific application domains only and there is a lack of generalisable architectural frameworks.
\end{itemize}

As an example, distributed AI systems combine properties of AI systems with properties of systems in IoT. This also means that challenges from both AI system design and IoT system design need to be accounted for when designing an architecture framework that can also support distributed AI systems. This article investigates whether a generalisable architecture framework is possible; the architecture framework should be able to cope with any number of challenges relevant for complex systems, such as AI systems in IoT.

A viable approach towards a generalisable architecture framework could be to utilise compositional thinking. Compositional thinking bases on category theory, which is a formalised way of representing structures and orders through directed graphs~\citep{Awodey2010}. \citet{Censi2017} proposed to use compositional thinking to ``co-design" functionality aspects of a complex system and the required resources such as hardware components. \citet{Zardini2020} demonstrated how applied category theory supports the co-design of hardware and software for an autonomous driving system, and \citet{Bakirtzis2021} applied compositional thinking to engineering of cyber-physical systems. \par

The main contribution is a mathematical description for compositional architectural frameworks. We first establish suitable descriptions of the abstractions levels for architectural views, their classification into clusters of concern, and elements from category theory to manage the relation between the views. Then, we show, based on an example, that the compositional approach to architectural frameworks can have the potential to ease co-design of systems that combine ``classical" system components, AI components, and concerns when integrating an AI system into the Internet of Things (IoT), and thereby ease the identified challenges. Finally, the proposed compositional architectural framework is demonstrated on a use case taken from a joint industry project.\par

\section{Research Method}

\begin{figure}[!b]
    \centering
    \includegraphics[width=0.6\linewidth]{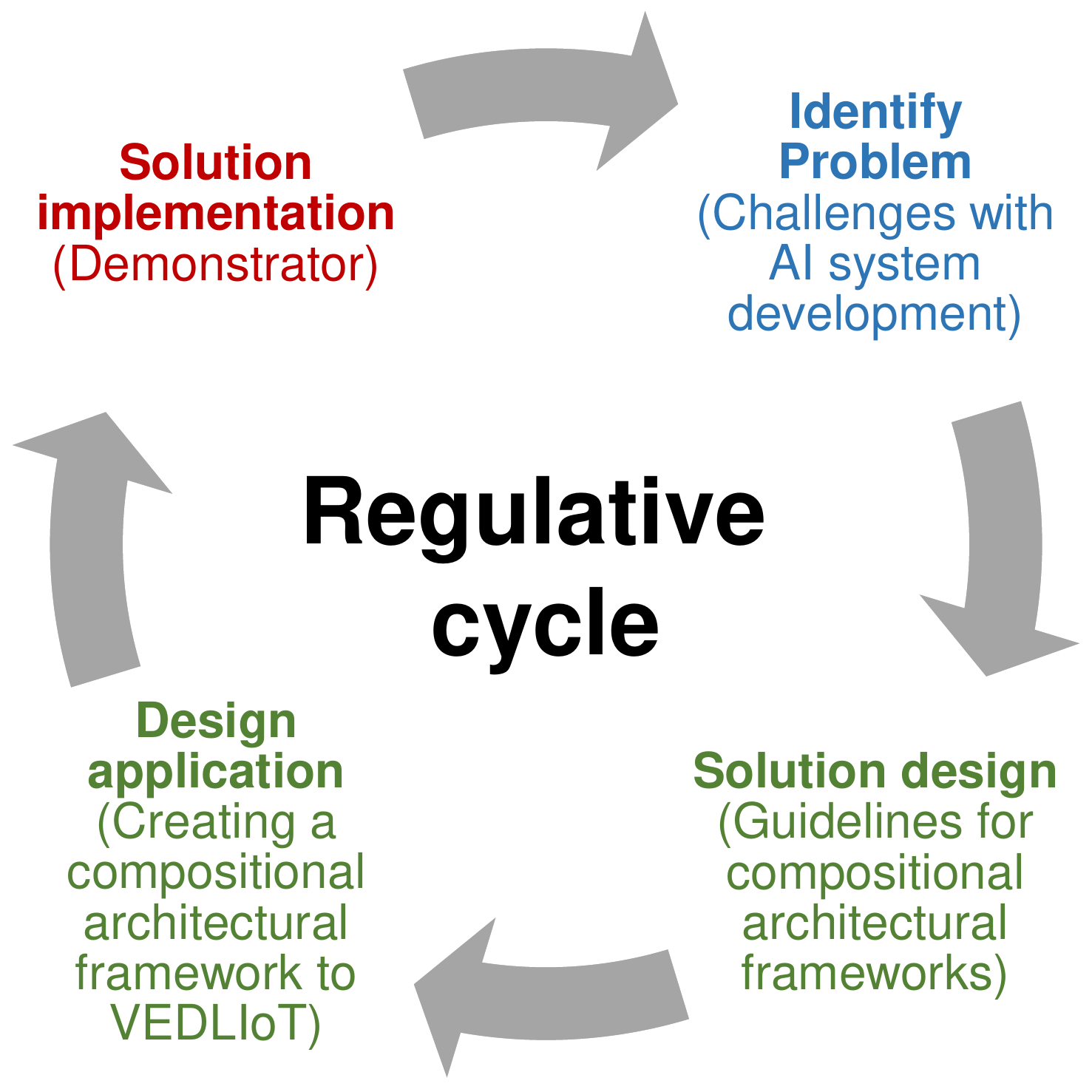}
    \caption{Regulative cycle of the design science process as proposed in \citet{Wieringa2009}.}
    \label{fig:02_cycle}
\end{figure}

The study follows the design-science paradigm, outlined, among others, in \citet{Hevner2004} and \citet{Peffers2007}. The study contains three major parts: (1) identification and motivation of the problem, (2) design, development, and validation of an artefact, and (3) demonstration of the artefact.
The different parts contribute to answering the following research questions:

\begin{description}
\item[RQ1:] Which challenges are relevant when defining system architectures for AI systems?
\item[RQ2:] What guidance can compositional thinking provide to overcome these challenges for the design of architectures for AI systems?
\item[RQ3:] How can a compositional framework be defined and applied in a realistic context?
\end{description}

\begin{figure}[!tb]
    \centering
    \includegraphics[width=\linewidth]{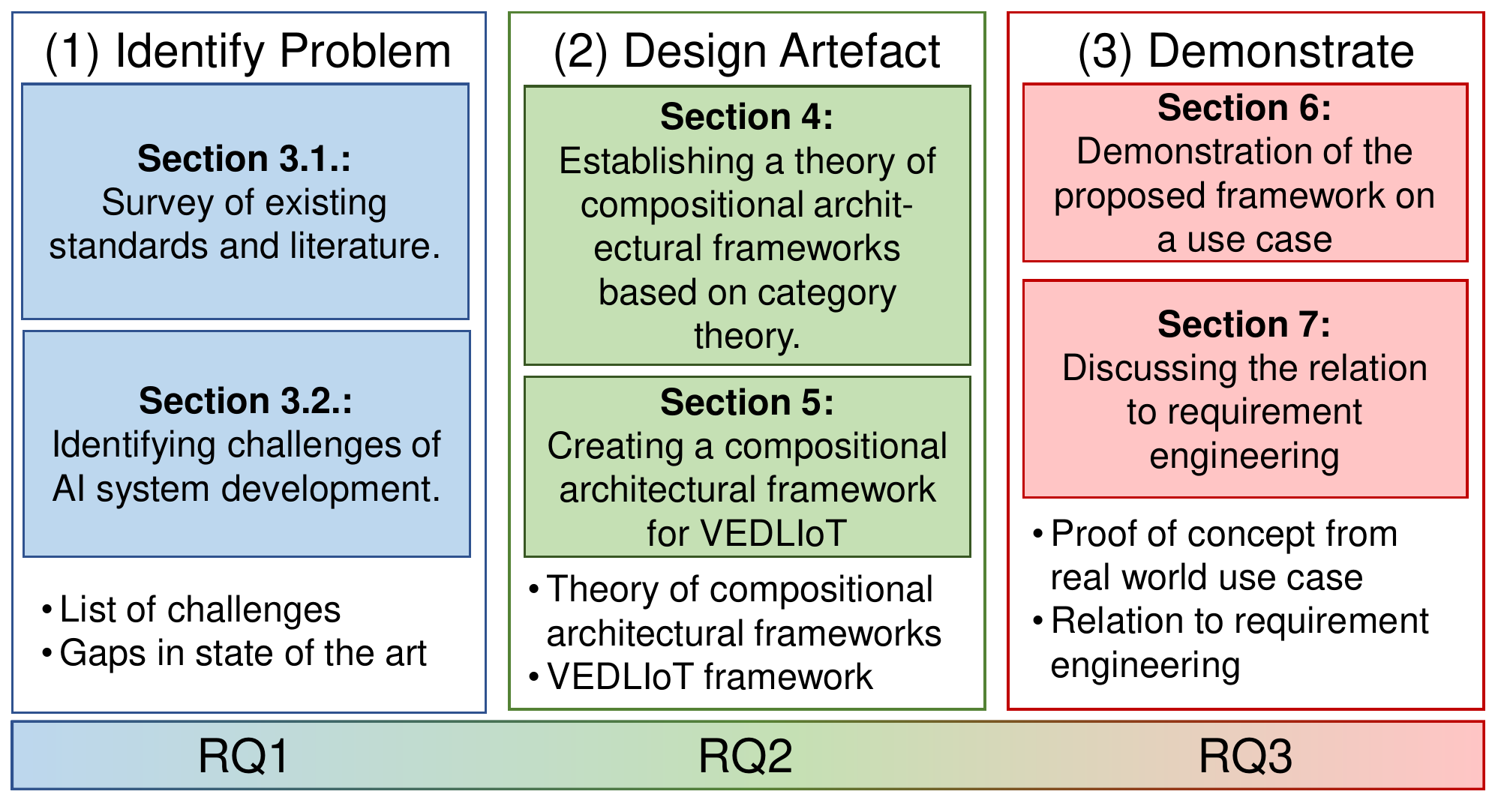}
    \caption{Structure of this study including the three parts of the research cycle.}
    \label{fig:02_structure}
\end{figure}

Following the concept \emph{design science as nested problem solving} proposed by \citet{Wieringa2009}, the different parts of the study represent a regulative cycle as illustrated in Figure~\ref{fig:02_cycle}. As reflected by the research questions, our intention was to investigate a problem, designed a solution, validated the solution by applying it to a realistic context, and finally implement the solution in form of a demonstrator. Figure~\ref{fig:02_structure} shows the resulting different elements of this study. \par
Through literature search and two focus groups with experts, Section~\ref{sec:03_challenges} produces a list challenges when developing architectures for distributed AI systems. Section~\ref{sec:04_framework} presents the artefact of the design science study, which encompasses the theory of compositional architectural frameworks. In Section~\ref{sec:05_application} the theory is tested by creating an architectural framework for the joint industry EU Horizon research project VEDLIoT. Following the design evaluation methods from \citet{Hevner2004}, this constitutes an observational case study. Section~\ref{sec:06_demonstration} shows the outcome of the solution implementation on a real world use case. According to \citet{Hevner2004}, this constitutes a descriptive evaluation using a detailed scenario around the artefact. Section~\ref{sec:07_discussion} establishes the relation between the architectural framework theory and requirements engineering.


\section{Challenges with architectural frameworks for complex systems}
\label{sec:03_challenges}
We first analysed relevant standards related to architectural frameworks with special attention to those specific for AI systems and IoT systems to derive the challenges with architectural frameworks for distributed AI systems. We complemented this analysis with a literature survey in systems engineering for AI to understand the current state of the art in research. We focused on AI and IoT systems because they represent systems of often high complexity that need to be supported by our proposed compositional approach to an architecture framework. We also conducted a workshop and focus groups with practitioners in order to truly understand and validate the challenges we need to account for in an architectural framework, especially towards our use case for distributed AI systems.

\subsection{Surveying existing standards and literature}
\label{subsec:03_LitChallenges}

Based on the architecture ontology and methodology of ISO 42010 \citep{ISO42010}, IEEE published a standard for architectural frameworks for IoT \citep{IEEE2413}. 
The purpose of this standard is to ``provide a framework for system designers to accelerate design, implementation, and deployment processes". 
Good overviews of standardisation attempts for IoT architectures are given by \citet{Weyrich2016}, \citet{Ray2018}, and in the architecture section of \citet{MohdAman2020}. \par
IoT is a network of cyber-physical devices and systems, and, although it does not directly address IoT, NATOs architecture framework provides an architectural framework for large distributed systems of intelligent agents \citep{NATO2020}.\par

\subsubsection*{Standardisation of architectural frameworks for AI}
In 2021, the international standardisation for architectural frameworks of AI systems is still ongoing. 
The only published international standard relevant for AI systems is currently ISO/IEC~TR~20547, which describes a standardisation of big data reference architectures~\citep{ISO20547}. 
Table~\ref{tab:03_AIStandards} provides an overview of ongoing international standardisation efforts. 

\begin{table}[!tb]
\centering
\caption{List of ongoing international standardisation related to architecture frameworks for AI system.}
\vspace{1em}
\label{tab:03_AIStandards}
\footnotesize
\begin{tabular}{lll}
\hline
\rowcolor[HTML]{C0C0C0}
\multicolumn{1}{c}{\textbf{Number}}              & \multicolumn{1}{c}{\textbf{Title}}                                                                                             & \multicolumn{1}{c}{\textbf{Status}}   \\       \hline                                  
\begin{tabular}[c]{@{}c@{}}ISO/IEC 5338\end{tabular}     & \begin{tabular}[l]{@{}l@{}} AI system lifecycle \\ processes\end{tabular}   &\begin{tabular}[l]{@{}l@{}} Preparatory \end{tabular} \\ 
\rowcolor[HTML]{EFEFEF}
\begin{tabular}[c]{@{}c@{}}ISO/IEC 5392\end{tabular}     & \begin{tabular}[l]{@{}l@{}}Reference architecture of \\ knowledge engineering\end{tabular}  &\begin{tabular}[l]{@{}l@{}} Preparatory\end{tabular}                                                  \\ 
\begin{tabular}[c]{@{}c@{}}ISO/IEC 5469 \end{tabular}& \begin{tabular}[l]{@{}l@{}}Functional safety\\ and AI systems    \end{tabular}             & \begin{tabular}[l]{@{}l@{}}Proposal\end{tabular}                                                     \\ \rowcolor[HTML]{EFEFEF}
\begin{tabular}[c]{@{}c@{}}ISO/IEC 23053   \end{tabular}& \begin{tabular}[l]{@{}l@{}}Framework for AI systems\\ using machine learning   \end{tabular}  & \begin{tabular}[l]{@{}l@{}}Under \\ Approval\end{tabular}    \\ 
\begin{tabular}[c]{@{}c@{}}ISO/IEC 24030    \end{tabular}& \begin{tabular}[l]{@{}l@{}}Artificial intelligence -\\ Use cases               \end{tabular}    & \begin{tabular}[l]{@{}l@{}}Under \\ publication\end{tabular} \\ \rowcolor[HTML]{EFEFEF}
\begin{tabular}[c]{@{}c@{}}ISO/IEC 24372  \end{tabular} & \begin{tabular}[l]{@{}l@{}}Overview of computational\\ approaches for AI systems \end{tabular}&\begin{tabular}[l]{@{}l@{}} In draft  \end{tabular}  \\ \hline      
\end{tabular}
\end{table}

\subsubsection*{State of the art for AI systems architecture}
In a research agenda for engineering AI systems, \citet{Bosch2020} provide a list of challenges when developing architectures for systems with AI components: 
(i) providing the right (quality of) data used for training, (ii) establishing the right learning infrastructure, (iii) building a sufficient storage and computing infrastructure, and (iv) creating a suitable deployment infrastructure. 
Indeed, the process of finding the data which provide a ground truth or reference for the AI to train on is often insufficiently documented during requirement engineering \citep{Kondermann2013}. 
This is insofar problematic as especially deep learning models depend on a vast amount of ground truth data for training and testing~\citep{Ries2021}.
Therefore, additional architectural views might be needed to capture the learning perspective of the AI part of the overall system \citep{Muccini2021}.
Because it might only be possible to detect and correct flaws in an AI systems after deployment, monitoring under operation of the system needs to become part of a suitable deployment infrastructure \citep{Bernardi2019}. 
Of course, an AI system does not only consist of AI components, but relies also on conventional software and hardware components. 
The development of AI components and traditional system components must therefore be aligned to avoid unwanted technical debt~\citep{Sculley2015}.

However, as Woods emphasises, traditional architecture frameworks, such as the 4+1 architectural view model by \citet{Kruchten1995}, does not account for data and algorithm concerns connected to AI component development~\citep{Woods2016}. 

There exist approaches to software architecture that treat data as explicit viewpoints, such as the one described in \citet[Chapter 2.6]{Clements2011}. These data viewpoints provide models and views for the data flow and data storage / management, which is relevant for distributed and connected systems, such as the IoT.

Context diagrams are an established method to capture the operational domain of the system \citep{Woods2009}, and architectural frameworks can consider them as own viewpoints \citep[Chapter 16]{Rozanski2012}. However, in a recent study we identified a number of challenges with context definitions for AI systems; for example, we found that they are often not in sync with the system development, and, therefore, often overly conservative, not well integrated into agile workflows, and disconnected from the function developers and other stakeholders \citep{Heyn2022}.

New stakeholders such as data engineers \citep{Vogelsang2019}, or governmental agency overseeing the use of AI in society~\citep{EuropeanCommission2020}) must be integrated in the system design and requirement engineering phases \citep{Altarturi2017}. 
A common platform needs to be found for communicating design decisions between requirement engineers, data engineers and other new stakeholders~\citep{Ahmad2021}.
The AI part can also causes new concerns to arise during system design, such as a stronger focus on ethical considerations \citep{Aydemir2018}, fairness or explainability.
Aspects such fairness \citep{Habibullah2021} and explainability \citep{Chazette2020} can be considered new non-functional requirements, which eventually require new architectural views describing them as quality aspects of the system \citep{Horkoff2019}. 
Developing AI components is a hierarchical, yet also iterative task: (i) prepare training data and/or environment, (ii) create a suitable model, (iii) train and evaluate the model, (iv) tune and repeat training, and finally, (v) deploy and monitor the run-time behaviour of the trained model~\citep{Bosch2020, Wan2020}. 
Its design needs to be decomposed into different levels of system design, and consistency needs to be ensured in order to satisfy high level requirements \citep{Giaimo2001}, and to fulfil the stakeholders' goals with a system. 
In addition, the system design must also allow for ``middle-out design", where existing components need to be integrated in the overall system design (e.g. transfer-learning from existing AI models or integration of off-the-shelf components)~\citep{Murugesan2019}. 
Murugesan et al. propose a hierarchical reference model, which supports the appropriate decomposition of requirements to the composition of the system's components. 
In their model the authors define how components can be decomposed into sub-components. 
To ensure consistency between the system architecture and the requirements, they define the terms \emph{consistency}, \emph{satisfaction}, and \emph{acceptability}. 
One major advantage of their model is that, if decomposition of system components is done correctly, these components can be independently specified and developed. 

\subsection{Identifying challenges of distributed AI system development in VEDLIoT} \label{subsec:03_Empiricalchallenges}

Two workshops with industry and academic partners from the VEDLIoT project\footnote{A brief description of the VEDLIoT project is given in~\ref{app:VEDLIoT}, and additional information can be found in~\citep{Heyn2021}.} were conducted with the aim of identifying and validating concerns relevant for a reference architecture framework for distributed AI systems from a practitioner's point of view. VEDLIoT is an excellent candidate for this study, because the aim of the project is to develop methods and tools for the development of distributed systems with deep learning components by using ``real world" use cases.

The first workshop took place in February 2021 and eleven participants joined the discussion through the remote conferencing software Zoom. A list of participants is provided in Table~\ref{tab:03_participants}.

\begin{table}[!tb]
\caption{Participants list for the workshop on challenges relevant for an architectural framework for VEDLIoT.}
\vspace{1em}
\label{tab:03_participants}
\centering
\footnotesize
\begin{tabular}{p{.5cm}p{3.5cm}p{1.4cm}p{1.6cm}}
\hline
\rowcolor[HTML]{C0C0C0}
\multicolumn{1}{c}{\textbf{No}} & \multicolumn{1}{c}{\textbf{Area of Expertise}}                                                          & \multicolumn{1}{c}{\begin{tabular}[c]{@{}c@{}}\textbf{Industry}\\ \textbf{Partner}\end{tabular}} & \multicolumn{1}{c}{\begin{tabular}[c]{@{}cl@{}}\textbf{Academic} \\ \textbf{partner}\end{tabular}} \\ \hline
1                      & Industrial IoT                                                             & \hspace{.7cm}\checkmark                                                      &                                                                                 \\ \rowcolor[HTML]{EFEFEF}
2                      & Smart Home                                                                 &                                                                                & \hspace{.7cm}\checkmark                                                       \\
3                      & Automotive Systems                                                         & \hspace{.7cm}\checkmark                                                      &                                                                                 \\ \rowcolor[HTML]{EFEFEF}
4                      & DL Optimization                                                            & \hspace{.7cm}\checkmark                                                      &                                                                                 \\
5                      & AI Hardware                                                                & \hspace{.7cm}\checkmark                                                      &                                                                                 \\ \rowcolor[HTML]{EFEFEF}
6                      & Requirement Engineering                                                    &                                                                                & \hspace{.7cm}\checkmark                                                       \\
7                      & IoT and AI research                                                        &                                                                                & \hspace{.7cm}\checkmark                                                       \\ \rowcolor[HTML]{EFEFEF}
8                      & AI systems development                                                     & \hspace{.7cm}\checkmark                                                      &                                                                                 \\
9                      & \begin{tabular}[c]{@{}l@{}}Secure conc. for IoT and AI\end{tabular} &                                                                                & \hspace{.7cm}\checkmark                                                       \\ \rowcolor[HTML]{EFEFEF}
10                     & AI Hardware Research                                                       &                                                                                & \hspace{.7cm}\checkmark                                                       \\
11                     & \begin{tabular}[c]{@{}l@{}}Systems Safety  Concepts\end{tabular}         &                                                                                & \hspace{.7cm}\checkmark                                                       \\ \hline
\end{tabular}
\end{table}

After explaining the aim of the workshop, the participants were presented with fundamental concepts of Architectural Frameworks for the IoT as described in IEEE~2413-2019~\citep{IEEE2413}. 
Table~1 of that standard provides a list of stakeholders for IoT systems which was distributed to the participants before the workshop.
After inspecting the table, the participants were asked, if additional stakeholders need to be considered when considering IoT systems with AI components.
The participants agreed that the list of common stakeholders from the standards contains most relevant stakeholders, and that additional stakeholder in regards to the AI components are data scientists and legislator and/or policy makers who might impose additional rules, e.g. for data privacy, transparency, or explainability of the AI's decisions. \par
In a second step, we wanted to identify relevant concerns for systems that are part of the IoT and, at the same time, contain AI components. 
The list of concerns for IoT systems given in Table~2 of IEEE~2413-2019~\cite{IEEE2413} was provided to the participants in advance of the workshop.  
During the workshop, the participants were asked to list all relevant concerns for IoT systems with AI components that are either on the standard's list of concerns, or that are not mentioned by the standard. \par
The results were collected in a mind-map and, together with the participants, clustered into what we will call ``clusters of concern". 
The resulted mind-map is reproduced in Figure~\ref{fig:03_ClustersofConcern} in \ref{app:mindmap} and can also be found in a repository holding a replication package and additional material\footnote{\url{https://doi.org/10.7910/DVN/VXFFFU}}.
Concern groups that are not already covered by IEEE~2413-2019~\citep{IEEE2413} have received an ID in the Figure and are summarised alphabetically in Table~\ref{tab:03_challenges}, together with the challenges identified from literature. The identified concerns were validated in a second workshop with the same participants in March 2021.

\begin{table}[!tb]
\caption{Summary of challenges identified from literature (L) and a workshop (W).}
\label{tab:03_challenges}
\centering
\footnotesize
\begin{tabular}{p{.33cm}p{3cm}p{.11cm}p{.12cm}p{3cm}}
\hline
\rowcolor[HTML]{C0C0C0} \textbf{ID} & \textbf{Description} & \textbf{L} & \textbf{W} & \textbf{Sources} \\\hline
\#1 & Additional views needed for describing the AI model & & \checkmark\\
\rowcolor[HTML]{EFEFEF} \#2 & Data requirements for ensuring the desired AI's behaviour must be considered & \checkmark        & \checkmark & \cite{Ries2021,Woods2016,Kondermann2013}\\
\#3 & Description of context and design domain & & \checkmark & \\
\rowcolor[HTML]{EFEFEF} \#4 & Describing the learning setting environment & \checkmark &            &  \cite{Muccini2021,Bosch2020,Woods2016}\\
\#5 & Integration of additional stakeholders (e.g., data scientists, policy makers) & \checkmark    &  & \cite{Ahmad2021,Vogelsang2019,Altarturi2017,Sculley2015}\\
\rowcolor[HTML]{EFEFEF} \#6 & Management of dependencies and correspondences between views & \checkmark &       & \cite{Nuseibeh2001}\\
\#7 & New quality aspects (e.g., explainability, fairness) & \checkmark &   \checkmark          & \cite{Habibullah2021,EuropeanCommission2020,Horkoff2019,Aydemir2018}\\
\rowcolor[HTML]{EFEFEF} \#8 & Run Time monitoring &  \checkmark        & \checkmark     & \cite{Wan2020,Bernardi2019}\\
\#9 & Support of decomposition into different levels of system design &     \checkmark    &  & \cite{NATO2020,Giaimo2001,Nuseibeh2001}\\
\rowcolor[HTML]{EFEFEF} \#10 & Support of middle-out design & \checkmark & & \cite{Murugesan2019}\\
\hline

\end{tabular}
\end{table}

\subsection{Problem statement}
From literature and the workshop we conclude that when combining architectural aspects for IoT and AI systems, as examples for the development of complex systems, many new concerns arise beyond traditional software engineering. 
Examples of the new concerns are data quality aspects, heuristic AI modelling, AI learning, and even ethical considerations. 
New stakeholder such as data engineers enter the stage, and common languages or interfaces need to be managed between the different stakeholders. 
Architectural views, governed through viewpoints, help to capture the different concerns from different stakeholders. 
However, typical architectural frameworks, such as the ISO~42010~\citep{ISO42010} or the IEEE~2413~\citep{IEEE2413} standard cannot cope with the large set of architectural views necessary to satisfy all stakeholders' concerns. One reason is, that certain architectural views were not foreseeable at time of creation of the standards, such as views relating to the explainablity or other ethical considerations of AI systems.
Another problem of current architectural frameworks is the lack of a clear system development hierarchy, which would support the early identification and mapping of dependencies between different architectural views~\citep{Nuseibeh2001}.\par
Finally, a major challenge we identified in the workshop is the difficulty to keep track of dependencies, e.g. through correspondence rules, between the different architectural views. Table~\ref{tab:03_challenges} lists the challenges identified for developing AI systems in IoT. In the following, we present our solution to cope with these challenges and in Section~\ref{sec:solutionChallenges} we discuss how the proposed solution mitigates the identified challenges.

\section{The concept of compositional architectural frameworks}  \label{sec:04_framework}
This section proposes the concept and artefact of a compositional architectural framework. In the regulative cycle shown in Figure \ref{fig:02_cycle}, this section describes the solution design.\par
The aim is not to provide a viewpoint catalogue or specific architectural models for AI systems, but instead to establish a structure for ensuring consistency and traceability of an architectural framework applicable to, for example, a distributed AI system.
Based on ideas from category theory and the identified challenges with architectural frameworks for AI and IoT systems, 
propositions are derived which provide ``rules" towards building a compositional architectural framework. 
Each proposition is clearly defined and demonstrated with a running example.
We call the framework \emph{compositional}, because it is built up from different ``modules", called clusters of concern, at different levels of abstraction.

\subsection{Compositional architectural framework theory} \label{subsec:03_Archiviews}
\begin{figure}[!htbp]
    \centering
    \includegraphics[width=\linewidth]{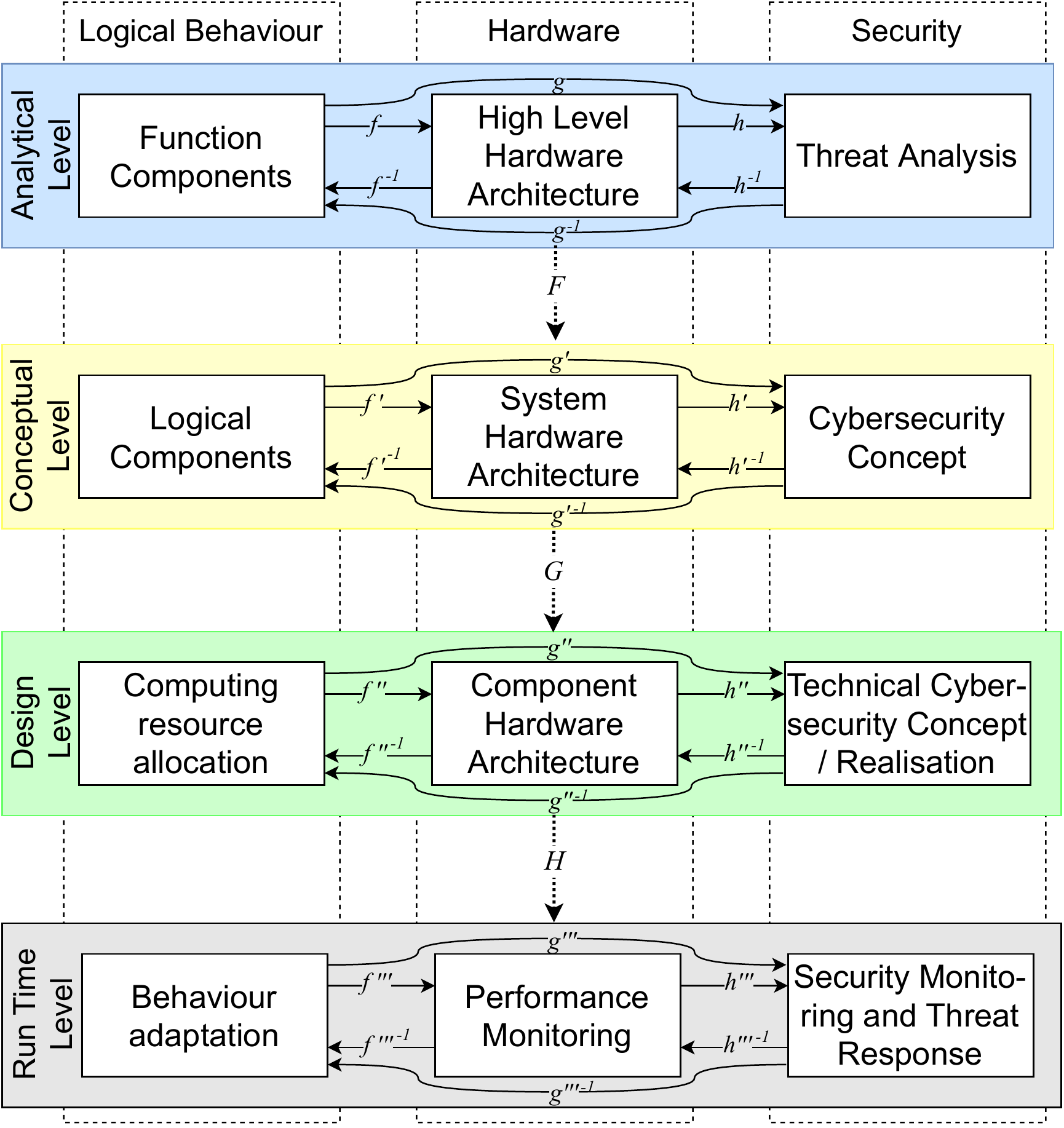}
    \caption{Order of Architecture Viewpoints}
    \label{fig:03_ArchViewpasorder}
\end{figure}

Figure \ref{fig:03_ArchViewpasorder} exemplifies three aspects of a system: logical behaviour, hardware, and cybersecurity. 
Each box represents a view governed by an architecture viewpoint on the final system. 
The rows represent levels of abstraction. For each aspects, the level of details increases with each additional level of abstractions. 
Let us call the levels, from top (highest level of abstraction) to bottom (highest level of details), analytical level, conceptual level, design level, and run time level. We chose these four ``default" levels of abstraction, because we think they represent typical system development phases.
However, depending on the type of systems and development regime, additional levels of abstraction can be added or removed, or different names for the levels can apply.\par


We show that by applying four rules, the architectural views for a system-of-interest can be arranged in a matrix, sorted by \emph{clusters-of-concern} and \emph{levels of abstraction}. During system development, different architectural views of the system-of-interest are created to describe different concerns with the system. 

\subsubsection{Cluster of concern}
We identified in Section~\ref{tab:03_challenges} that for complex systems, such as for example distributed AI systems, many different concerns need to be considered in the final system's architecture. Clustering the concerns into sets is a starting point towards the architectural framework:

\begin{tcolorbox}[title=Cluster of concern]
\small
\textit{A cluster of concern is a partially ordered set of architectural views which represent a specific concern of the system at different levels of details.}
\end{tcolorbox}
\begin{definition}[Cluster of concern]
\label{prop0}
Let $A$, $B$, and $C$ be architectural views forming a set $X$ called cluster of concern. Assume, that $B$ conveys more or equal details about the system than $A$, and $C$ conveys more or equal details than $B$. We define a binary relation $\leq$, such that $A \leq B$. It further holds that: 
\begin{enumerate}
    \item $A \leq A$ (reflexivity);
    \item if $A \leq B$ and $B \leq C$, then $A \leq C$ (transitivity);
    \item if $A \leq B$ equals $B \leq A$, then $A$ and $B$ are of equal level of detail (equivalent) and we write $A = B$.
\end{enumerate} 
The pair $(X,\leq)$ is then a partially ordered relation.
\end{definition}

\begin{example*}
The architectural view ``computing resource allocation" contains more details about the final system than the architectural view ``logical components". The architectural view ``function components" contains the least amount of details. These three architectural views form an ordered set of architectural views considering the concern ``logical behaviour". Therefore, they form the cluster of concern ``Logical Behaviour".
\end{example*}

\subsubsection{Levels of abstraction}
The further the system design proceeds, the more details about the final system become apparent. This increase in level of details during system development allows for defining levels of abstractions. The views are ordered in the sense that the level of detail increases with each level of abstraction:

\begin{tcolorbox}[title=Level of abstraction]
\small
\textit{The set of architectural views with equivalent level of details about the system-of-interest constitutes a category. A category is a collection of objects which are related to each other in a consistent way \citep{Perrone2021}. We call the category \emph{level of abstraction}. Architectural views on a level of abstraction are related to each other through morphisms. A morphism can exist only between architectural views on the same level of abstraction.}
\end{tcolorbox} 


\begin{definition}[Level of abstraction]
\label{prop1}
The category $\mathbf{C_{la}}$ describing a level of abstraction has architectural views as its objects. 
Given two architectural views $A$ and $B$, then $f\in\text{Hom}_\text{la}(A,B): A \rightarrow B$ describes the morphisms between architectural views of different concerns at the same level of abstraction. 
The identity morphism is the identify function mapping an architectural view to itself. 
Furthermore, given an additional architectural view $C$, and defining the additional morphism $h\in \text{Hom}_\text{la}(B,C): B \rightarrow C$\footnote{The morphism is called $h$ here instead of the expected $g$ to ensure consistency with later definitions}, there exists a morphism $f \fatsemi h$ such that the composition of $f$ and $h$ is $g \in \text{Hom}_\text{la}(A,C): A \rightarrow C$.
\end{definition}

\begin{example*}
In the example shown in Figure \ref{fig:03_ArchViewpasorder}, a morphism is the correspondence between the logical components and the system hardware architecture on the concept level.
This relation can also exist in both directions: The architecture of the system hardware architecture can correspond to the logical components. 
The morphisms between views that correspond to each other are therefore \textit{isomorphism}, i.e. $f \fatsemi f^{-1} = \text{id}$, where $\text{id}$ is the identity relation. 
\end{example*}

\subsubsection{Consistency of an architecture}
Describing an architecture through category theory yields the advantage that one can use mathematical tools to show and proof relations between the elements of an architecture description. For example, the product of two objects $X$ and $Y$ in category theory describe ``the most efficient way" to have both $X$ and $Y$. This can be utilised for finding rules on how to combine architectural views into a consistent system architecture description:

\begin{tcolorbox}[title=Consistency of architectures] 
\small
\textit{If the product over all architectural views on a level of abstraction is valid\footnote{Valid means, informally, that no matter which architectural view one ``starts at", it is always possible to ``transit" via the correspondence rules to another architectural view, and still see the same system (from a different perspective)}, the architectural views consistently describe the system-of-interest.} 
\end{tcolorbox}

\begin{definition}[Consistency of architectures]
\label{prop2}
Let $\mathbf{C_{la}}$ be a category and let $A$ and $B$ be two architectural views as objects of $\mathbf{C_{la}}$. 
The product of $A$ and $B$ consists of a new object $Z$ (this is the product), and two morphisms $\pi_1: Z \rightarrow A$ and $\pi_2: Z \rightarrow B$. The product is valid if the following diagrams commutes\footnote{Successfully commuting between different views means to ``look" at the system with different architectural views, without encountering inconsistencies in the system.}:
\par
\noindent
\center
\includegraphics[width=0.4\linewidth]{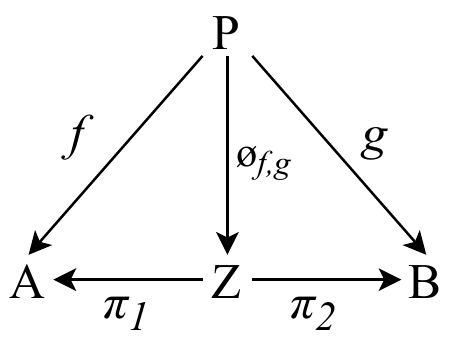}
\end{definition}

\begin{example*} 
Two architectural views $A$ and $B$, for example a system hardware architecture and a cybersecurity concept, can be combined to a new view $Z = A \times B$ which unites, in ``the most efficient way", the hardware architecture and the cybersecurity concept. 
$P$ now is any other view, for example the logical components.
In addition to the already existing relations $f$ and $g$ between the views, there must be a relation $\phi_{f,g}$ from the logical components $P$ to the newly combined ``secure hardware architecture" $Z$. 
Furthermore, there must be relations $\pi_1$ and $\pi_2$, such that one can go always from the ``secure hardware architecture" $Z$ back to the origin system hardware architecture $A$, or to the origin cybersecurity concept~$B$. \par
The product is valid, if, and only if, it does not matter if one first uses correspondence rules to go from the logical components to the new ``secure hardware architecture", and from there to the origin system hardware architecture ($\phi_{f,g} \fatsemi \pi_1$); or if one commutes from the logical components directly to the system hardware architecture ($f$). 
If it is not possible to commute between views at the same level of abstraction, one or several views are not describing completely all necessary aspects of the system-of-interest. 
\par
\end{example*}

\subsubsection{Mapping of relations}
Each level of abstraction is another category but Definitions \ref{prop1} and \ref{prop2} only describe relations between architectural views of the same category (i.e., on the same level of abstraction). However, another concept from category theory called functors allow for mappings between different categories (i.e., different levels of abstraction) that preserve and respect the relations between the objects \citep{Perrone2021}:

\begin{tcolorbox}[title=Mapping of relations]
\small
\textit{Functors map all views of one level of abstraction to corresponding views of the next lower level of abstraction, and all relations between views to corresponding relations of the next lower level of abstraction.}
\end{tcolorbox} 

\begin{definition}[Mapping of relations]
\label{prop3}
Let $\mathbf{C_{la}}$ and $\mathbf{C_{lb}}$ be categories representing two levels of abstractions in an architectural description. 
A functor $F$ from $\mathbf{C_{la}}$ to $\mathbf{C_{lb}}$ maps each object of $\mathbf{C_{la}}$ to a corresponding object of $\mathbf{C_{lb}}$; and maps each morphism between the objects in $\mathbf{C_{la}}$ to corresponding morphisms in $\mathbf{C_{lb}}$. 
Furthermore, unitality holds, i.e. identities $\text{id}$ are mapped into identities, and compositionality holds, i.e. let $f$ and $g$ be morphisms, then $F(f\fatsemi g) = F(f) \fatsemi F(g)$.
\end{definition}

\begin{example*}
If there exists a relation (morphism) between the Cybersecurity Concept and the System Hardware Architecture on the conceptual level, a corresponding relation (morphism) must exists on the design level between the technical cybersecurity concept and the component hardware architecture. 
Note, however, that the reverse is not necessarily true. If there exists a relation between two views on a lower level of abstraction, this relation does not necessarily exist on a higher abstraction level too. 
\end{example*}

\begin{figure}[htb]
    \centering
    \includegraphics[width=\linewidth]{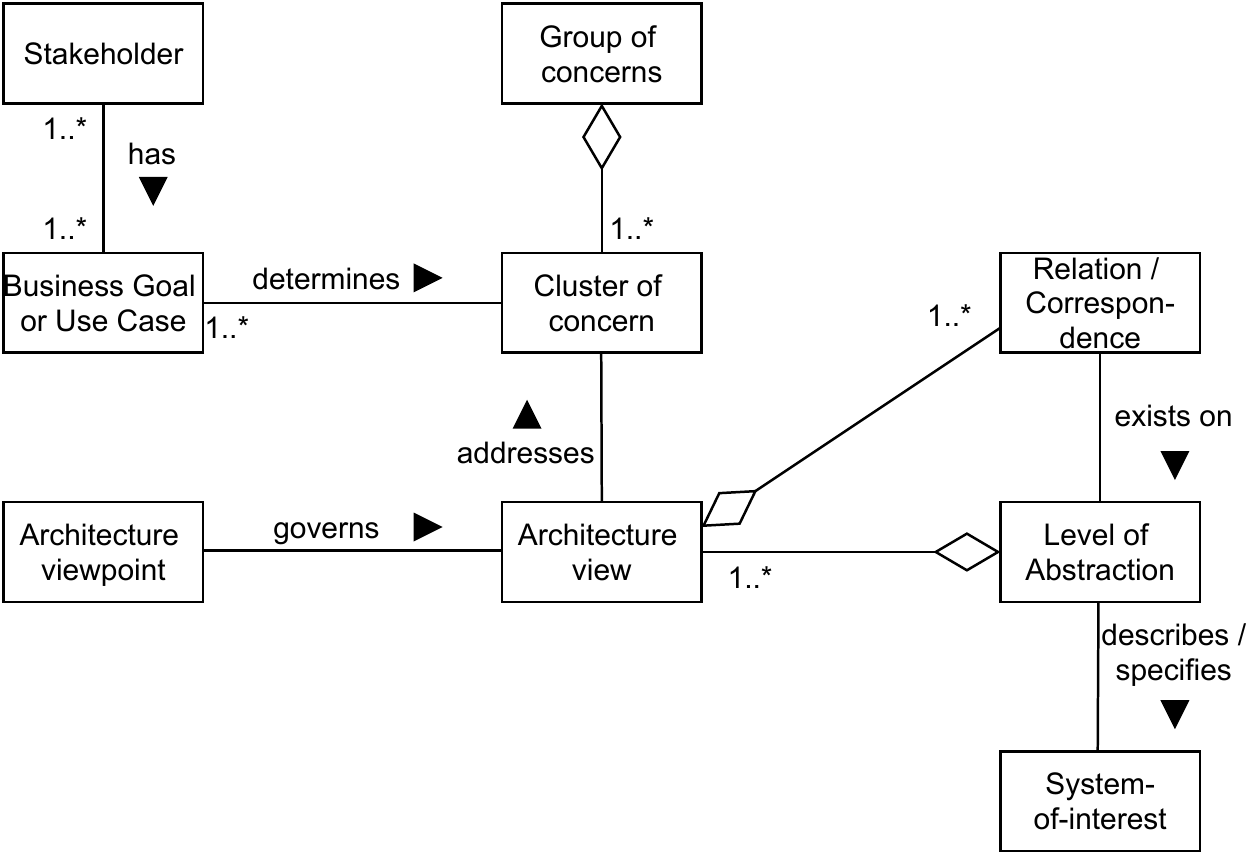}
    \caption{Conceptual model of a compositional architecture framework}
    \label{fig:03_ArchiFramework_UML}
\end{figure}

The final conceptual model of a compositional architecture framework based on the stated definitions is illustrated in Figure \ref{fig:03_ArchiFramework_UML}. 
Note that an additional object called ``Group of concerns" has been introduced. It serves as a way to order the clusters of concern into different major aspects of the system.

\subsection{Connecting the theory to the challenges}\label{sec:solutionChallenges}
Table~\ref{tab:03_challenges} listed the challenges identified for developing AI systems in IoT. The idea of a compositional architectural framework can solve these challenges as follows:

{\small 
\begin{itemize}[leftmargin=4mm]
\item {\em \#1: Additional views needed for describing the AI model} - 
\noindent Architectural views describing e.g., the configuration and hyper-parameter settings of the AI model can be explicitly taken into consideration through an own cluster of concern.
\item {\em \#2: Data requirements must be considered} - 
\noindent Depending on what the data concerns, data aspects can be integrated into views of different clusters of concern (e.g., training data can be handled as an own cluster of concern with views representing the data strategy for AI training). Data concerns regarding communication and information can be separated into another cluster of concern with own architectural views.
\item {\em \#3: Description of context and design domain} - \noindent The context and operation design domain can be treated as own clusters of concern, thus making the treatment of the context and design domain explicit during system development.
\item  {\em \#4: Describing the learning setting / environment} - 
\noindent The learning setting can be integrated as an own cluster of concern, with independent architectural views. The views can include objectives of the learning, learning scenarios selection, and a view describing the specific learning settings.
\item {\em \#5: Integration of additional stakeholders } - 
\noindent The proposed architectural framework is scalable in the sense that additional clusters of concern can easily be added. This allows for additional stakeholders to add their concerns and (architectural) views for the system in own clusters of concern.
\item {\em \#6: Management of dependencies and correspondences between views} - 
\noindent A compositional architectural framework provides rules on how to handle dependencies between architectural views. The existence of correspondence rules in a compositional architecture is limited by definitions~\ref{prop1} and~\ref{prop3}. Specifically, correspondence rules should only exist between architectural views on the same level of abstraction. 
\item {\em \#7: New quality aspects (e.g., explainability, fairness)} - 
\noindent New (quality) concerns such as explainability can be explicitly integrated in the architectural framework as clusters of concern. The concept of compositionality allows for an easy scalability, because any number of additional clusters of concern can be added to the framework.
\item {\em \#8: Run time monitoring} - 
\noindent Run time concerns, such as run time monitoring, can be made explicit with an own level of abstraction, as suggested in the default setup of a compositional architectural framework.
\item {\em \#9: Support of decomposition into different levels of system design} - 
\noindent A compositional architectural framework explicitly supports the decomposition into different levels of system design. The number of levels of system design is flexible, and can be adapted to the needs of the system and company.
\item {\em \#10: Support of middle-out design} - 
\noindent Middle-out design is supported because it is not required to ``fill" the framework with architectural views from top to bottom. For example, it is possible to define a component hardware architecture first (because a certain hardware component must be used in the system), and then build the system hardware architecture ``around it". Definition \ref{prop2} can be used to ensure that no inconsistencies between architectural views on the same level of abstraction exist after finishing the middle-out design.
\end{itemize}
}

\section{A compositional architecture framework for VEDLIoT} \label{sec:05_application}
The idea of a compositional architectural framework is tested by studying the artefact in depth in the VEDLIoT project. The VEDLIoT project is a suitable test-ground because it combines conventional systems with deep learning and the IoT. This requires a variety of design decisions with different architecture views on the resulting systems. Additionally, VEDLIoT aims at explicitly supporting all necessary non-functional aspects of the system, such as data privacy, safety, but also non-functional aspects particularly relevant to deep learning such as explainability. The success criteria of the study were that the proposed framework should result in an applicable architecture framework for VEDLIoT, and that guidelines are found on how the theoretical framework, described in Section~\ref{sec:04_framework}, can be applied in a practical setting.

\subsection{Methodology of the case study}

\begin{figure}[!hbtp]
    \centering
    \includegraphics[width=\linewidth]{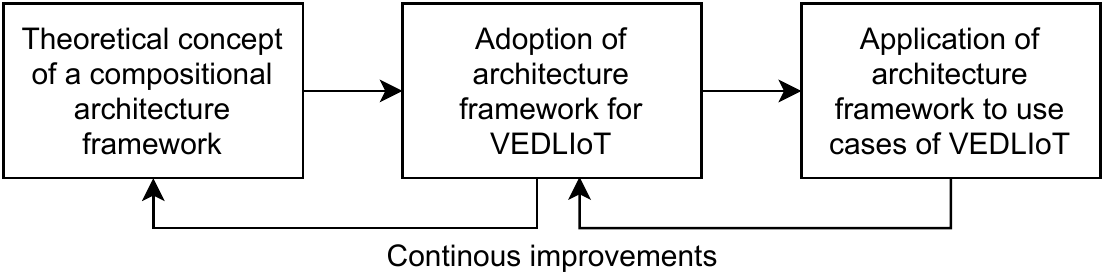}
    \caption{Iterative process of adopting and applying the compositional architecture framework concept to VEDLIoT.}
    \label{fig:04_ProcessFramework}
\end{figure}


\begin{table}[!b]
\caption{Participating roles in the development of the compositional architectural framework for VEDLIoT 
}
\label{tab:04_participants}
\centering
\footnotesize
\begin{tabular}{clccc}
\hline
\rowcolor[HTML]{C0C0C0}
\multicolumn{1}{c}{\textbf{No}} &  \multicolumn{1}{c}{\textbf{Role}}                                                          & \multicolumn{1}{c}{\begin{tabular}[c]{@{}c@{}}\textbf{Industry}\\ \textbf{partner}\end{tabular}} & \multicolumn{1}{c}{\begin{tabular}[c]{@{}c@{}}\textbf{Academic} \\ \textbf{partner}\end{tabular}} &
\multicolumn{1}{c}{\begin{tabular}[c]{@{}c@{}}\textbf{Years of} \\ \textbf{experience}\end{tabular}}\\ \hline
1                      & Researcher &   &   \checkmark          & 11     \\ \rowcolor[HTML]{EFEFEF}
2                      &  \begin{tabular}[l]{@{}l@{}}Research\\ Engineer\end{tabular} &  \checkmark        &     & 4 \\
3                      & Researcher &  \checkmark        &      & 20+   \\\rowcolor[HTML]{EFEFEF}
4                      & Researcher &    &   \checkmark          & 30           \\
5                      & Developer &     \checkmark    &    & 1               \\\rowcolor[HTML]{EFEFEF}
6                      & PhD Student &         &     \checkmark   & 3        \\
7                      & Professor &                &  \checkmark      & 15        \\\rowcolor[HTML]{EFEFEF}
8                      & \begin{tabular}[l]{@{}l@{}}Deep Learning \\ Developer \end{tabular} &    \checkmark                &        & 3  \\
9                      & Developer &     \checkmark               &      & 5     \\\rowcolor[HTML]{EFEFEF}
10                     & Project Lead &  \checkmark               &          & 15    \\
11                     & Researcher &         &  \checkmark       & 1      \\ \rowcolor[HTML]{EFEFEF}
12                     & Researcher &         &  \checkmark & 4       \\
13                     & Researcher &       &  \checkmark & 2        \\ \rowcolor[HTML]{EFEFEF}
14                     & Professor &          &  \checkmark        & 20+     \\
15                     & Professor &          &  \checkmark       & 6  \\ \rowcolor[HTML]{EFEFEF}
16                     & Developer &     \checkmark      &        & 10     \\
\hline
\end{tabular}
\end{table}

Participants from all involved companies and academic partners met in bi-weekly meetings to analyse the use cases, to adopt the architectural framework to the needs of the VEDLIoT project, and ultimately to apply the compositional architectural framework to the use cases.
Figure~\ref{fig:04_ProcessFramework} shows the iterative character of this process: Initially, the participants were introduced to a preliminary version of the theoretical idea of a compositional architectural framework. 
A first step was to find a general adoption of the architecture framework for VEDLIoT, which was then continuously applied to the use cases.
If the theoretical framework or the general adoption did not fit to the use cases, changes in the theoretical concept of the architectural framework would be proposed and implemented.
As is common practice in workshops related to solution design\footnote{see \citet[Chapter 10]{Hevner2010}}, the authors of this article were participants in the meetings, and therefore influenced the development of the architectural framework for VEDLIoT.
However, the majority of participants were not involved in the academic observations, and therefore could validate or rebut the ideas brought forward by the authors. 
A list of the roles and experience of workgroup's participants is given in Table~\ref{tab:04_participants}.
The entire version history and evolution of the VEDLIoT architectural framework can be found in the supplement material\footnote{\url{https://doi.org/10.7910/DVN/VXFFFU} archi framework}.



\subsection{Clusters of concern} \label{04_subsec:Concerns}


The challenges of architectural frameworks for distributed AI systems, summarised in Table~\ref{tab:03_challenges}, were used as starting point for discussions on the necessary clusters of concern during bi-weekly meetings of the VEDLIoT partners. 
First, the group identified four major groups of concerns for the architecture framework. 
Then, for each of the groups of concern, the participants of the bi-weekly meetings analysed the use cases of VEDLIoT and determined the required clusters of concern. 
The overall aim was to create as many clusters of concern as the VEDLIoT use cases require (none missing), yet trying to minimise the amount of clusters of concern (no cluster of concern can be removed). 
The resulting clusters of concern are summarised in Table~\ref{05:tab_VEDLIoTClusters}. Note, that other development project might require different clusters of concerns. The decision which clusters are required is based on the use case and business goals, as highlighted in the top row of the framework.

\begin{table}[!htb]
\centering
\caption{Description of clusters of concern in the VEDLIoT framework. Dark: Groups of concerns.}
\label{05:tab_VEDLIoTClusters}
\begin{singlespace}
\footnotesize
\begin{tabular}{cl}
\hline
\rowcolor[HTML]{C0C0C0} 
\multicolumn{1}{c}{\cellcolor[HTML]{C0C0C0}\textbf{Concern}} &
  \multicolumn{1}{c}{\textbf{Description}} \\ \hline
  \cellcolor[HTML]{d7d7d7}\begin{tabular}[c]{@{}c@{}}Behaviour \\and Context\end{tabular} &
  \cellcolor[HTML]{d7d7d7}\begin{tabular}[c]{@{}l@{}}Aspects that concern the static and dynamic\\ behaviour of the system, as well as the context\\ and constraints for the desired behaviour.\end{tabular} \\
  \hline

  \begin{tabular}[c]{@{}c@{}}Logical\\ Behaviour\end{tabular} &
  \begin{tabular}[c]{@{}l@{}}Views that are concerned with the\\ static behaviour of the system.\end{tabular} \\
  
  \cellcolor[HTML]{EFEFEF}\begin{tabular}[c]{@{}c@{}}Process \\ Behaviour\end{tabular} &
  \cellcolor[HTML]{EFEFEF}\begin{tabular}[c]{@{}l@{}}Views concerned with the\\ dynamic behaviour of the system.\end{tabular} \\
  
  \begin{tabular}[c]{@{}c@{}}Context and \\ Constraints \end{tabular} &
  \begin{tabular}[c]{@{}l@{}}Contains views on the system that define\\ the context and limit the design domain. \end{tabular} \\
  \hline
  \cellcolor[HTML]{d7d7d7}\begin{tabular}[c]{@{}c@{}}Means and\\ Resource\end{tabular} &
  \cellcolor[HTML]{d7d7d7}\begin{tabular}[c]{@{}l@{}}Contains views on aspects of the system\\ that enable the desired behaviour.\end{tabular} \\
  \hline
    
  \begin{tabular}[c]{@{}c@{}}Hardware\end{tabular} &
  \begin{tabular}[c]{@{}l@{}}Includes views on the hardware architecture\\ and component design of the system. \end{tabular} \\

  \cellcolor[HTML]{EFEFEF}\begin{tabular}[c]{@{}c@{}}AI models \end{tabular} &
  \cellcolor[HTML]{EFEFEF}\begin{tabular}[c]{@{}l@{}}Contains views that describe the setup and\\ configuration of the required AI model.\\ Views can include model design, e.g.,\\neural network setup or views detailing\\ the configuration of the AI model. \end{tabular} \\
  
  \begin{tabular}[c]{@{}c@{}}Data strategy\end{tabular} &
  \begin{tabular}[c]{@{}l@{}}Views that support collection and selection for\\ training, validation, and run time data of the\\ AI model. Views can describe methods for data\\ creation, data selection, data preparations, and\\ run time monitors of data used by the AI. \end{tabular} \\
  
  \cellcolor[HTML]{EFEFEF} \begin{tabular}[c]{@{}c@{}}Learning\end{tabular} &
  \cellcolor[HTML]{EFEFEF}\begin{tabular}[c]{@{}l@{}}Covers views on the system that allow for de-\\ fining and setting up the learning environ-\\ ment of the AI model. This can include the\\ definition of training objectives and views that\\ outline the chosen optimiser for training. \end{tabular} \\
  \hline
  
  \cellcolor[HTML]{d7d7d7}\begin{tabular}[c]{@{}c@{}}Comm-\\unication\end{tabular} &
  \cellcolor[HTML]{d7d7d7}\begin{tabular}[c]{@{}l@{}}Contains views of data, connectivity and\\ communication between nodes or components\\ of the desired system.\end{tabular} \\
  \hline
  
  \begin{tabular}[c]{@{}c@{}}Information \end{tabular} &
  \begin{tabular}[c]{@{}l@{}}Accumulates views on the system that\\ model the information and data exchanged\\ in and through the system-of-interest. \end{tabular} \\
  
  \cellcolor[HTML]{EFEFEF}\begin{tabular}[c]{@{}c@{}}Connectivity\end{tabular} &
  \cellcolor[HTML]{EFEFEF}\begin{tabular}[c]{@{}l@{}}Contains views on the means of communication\\ available to the system and its resources. \end{tabular} \\
  
  \hline
  \cellcolor[HTML]{d7d7d7}\begin{tabular}[c]{@{}c@{}}Quality\\Concerns\end{tabular} &
  \cellcolor[HTML]{d7d7d7}\begin{tabular}[c]{@{}l@{}}Encompass quality aspects which can be\\ described through non-functional requirements\\ which affect the architecture of the system.\end{tabular} \\
  \hline

  \begin{tabular}[c]{@{}c@{}}Ethics \end{tabular} &
  \begin{tabular}[c]{@{}l@{}}Views that regulate ethical aspects, such as\\ fairness or transparency of the system. \end{tabular} \\

  \cellcolor[HTML]{EFEFEF}\begin{tabular}[c]{@{}c@{}}Security \end{tabular} &
  \cellcolor[HTML]{EFEFEF}\begin{tabular}[c]{@{}l@{}}Views that ensure the security\\ aspects of the system.  \end{tabular} \\
  
  \begin{tabular}[c]{@{}c@{}}Safety \end{tabular} &
  \begin{tabular}[c]{@{}l@{}}Contains views governing the safety\\ aspects of the system. The views can\\ stem from standards such as ISO~26262. \end{tabular} \\

  \cellcolor[HTML]{EFEFEF}\begin{tabular}[c]{@{}c@{}}Energy\\Efficiency \end{tabular} &
  \cellcolor[HTML]{EFEFEF}\begin{tabular}[c]{@{}l@{}}This cluster of concern contains views\\ ensuring energy efficiency,\\ especially for mobile devices. \end{tabular} \\
  
  \begin{tabular}[c]{@{}c@{}}Privacy \end{tabular} &
  \begin{tabular}[c]{@{}l@{}}Here views can be contained that ensure\\ privacy requirements, such as for example\\ requested by regulatory authorities. \end{tabular} \\
  
  \hline
\end{tabular}
\end{singlespace}
\end{table}

Some of the clusters of concern are especially relevant towards distributed AI systems. For example, the \textit{Context and Constraints} cluster of concern covers views on the system that define the context and limits the design domain. 
An example of a context viewpoint is described in \citet{Rozanski2012}. 
For AI systems, it is beneficial, sometimes even required, to explicitly state the desired context and to define views on the constraints and the (operational) design domain of the system. 
\citep{Ali2010} for example state, that the desired context is often ignored when defining requirements for a system, and \citep{Berry2022} underlines the need of context information for the assessment of the AI system's performance. 
In addition, Knauss et al. argue that run time uncertainty can be removed by making the context, in which requirements are valid, explicit~\citep{Knauss2016}. 
The context, in which the system operates, will influence architectural decisions (and vice versa), and thus should be made explicit during the design process. 

For distributed AI systems, the concern \textit{AI models} is relevant, because it contains views that describe the setup and configuration of the required AI models. 
Classification of objects in an optical videostream for example requires a different deep neural network setup then recognising natural language or predicting trajectories of other vehicles. 
Choosing the right AI model is a design decision which requires suitable views on the AI model in relation to the overall system. 
Also, the learning strategy of the AI model has paramount impact on the final behaviour of the AI system. 
Trained with a flawed datasets (e.g. biased data), the behaviour of the AI system will exhibit the flaws learned during the learning process (e.g., the trained system will exhibit a bias). 
The learning process for the AI model is therefore integral part of the system design process. For VEDLIoT we decided to describe the learning process in two clusters of concerns: The first learning specific cluster of concern, titled \textit{Data Strategy}, contains views that allow for the specification of the collection and preparations of the required training, testing, and run time data. The second learning-specific cluster of concern is titled \textit{Learning}. It covers views that allow for the definition of the learning environment, for example views that describe what the AI model is supposed to learn, and how it can learn, i.e., it can contain views elaborating the optimiser and learning settings for the training phase.
The concerns of AI model and learning have many dependencies between each other, which will be expressed through correspondences (morphisms).\par
Unlike previous architectural frameworks for the IoT, such as IEEE~2413-2019 \citet{IEEE2413}, the compositional thinking in the architectural framework allows for co-designing the system to fulfil the explicitly identified quality concerns, such as safety, security, but also energy efficiency and ethical concerns. 
It means that already early in the system development, correspondences between the views regarding the quality concerns and other views in the architecture description are established. 
The final system can then be said to be ``Safe by design", ``Secure by design", ``Efficient by design", or ``Fair by design". Recent legislation shows that the ethical aspects become a central concern when developing AI systems \citep{EuropeanCommission2020}.

\begin{table*}[!hbtp]
\centering
\caption{Description of the levels of abstraction (LoA).} 
\label{05:tab_VEDLIoTLevels}
\footnotesize
\begin{tabularx}{\linewidth}{cX}
\hline
\rowcolor[HTML]{C0C0C0} 
\multicolumn{1}{c}{\cellcolor[HTML]{C0C0C0}\textbf{LoA}} &
  \multicolumn{1}{c}{\textbf{Description}} \\ \hline

  Analytical level &
  The first level of abstraction includes architectural views that provide an abstract and high level view on the system-of-interest.
On that level, all views provide a way to describe the system and context on a knowledge level, which provides information for further, more concrete system development. For example, the high level AI model view could elaborate on which functions should be fulfilled through an AI.\\

  \cellcolor[HTML]{EFEFEF}Conceptual level &
  \cellcolor[HTML]{EFEFEF}On the next level of abstraction, the views provide a more concrete description of the overall system-of-interest.
Components are not detailed yet, but the overall system composition becomes clear and the context of operation is clearly defined. 
For example, the AI model could be concretely shaped as a Deep Learning Network with the required amount of layers. 
All views on this level combined provide a system specification that sets the system-of-interest in context and elaborates on how the desired functionality is fulfilled. \\

  Design level & 
The most concrete level at design time of the system is the design level, which includes views that concretely shape the final system-of-interest. 
Resources are allocated to components, the AI model is configured to work most efficiently in the given environment, and the concrete component hardware architecture is defined. 
The solution specification describes the final embodiment of the system-of-interest.\\

  \cellcolor[HTML]{EFEFEF}Run time level &
\cellcolor[HTML]{EFEFEF}Complex systems, both AI driven and conventional, often require forms of monitoring and operations control. 
The purpose of the run time monitoring can be manifold: On one hand, monitoring of a deployed system at run time provides valuable feedback about its performance and reliability to developers and product owners. 
DevOps is an essential component of an agile development framework, and early detection of issues in a deployed system allows for a swift response from the developers.\\
  \hline
\end{tabularx}
\end{table*}

\subsection{Levels of abstraction}
The architectural views are not only sorted by clusters of concern but also by their represented level of abstraction, as was discussed in Section \ref{subsec:03_Archiviews}. 

For VEDLIoT, the workgroup decided to follow the four proposed default levels of abstraction introduced in Section~\ref{sec:04_framework}. 
They are the \emph{analytical level} providing a high level view, the \emph{conceptual level} providing a more concrete but not too detailed description of the system, the \emph{design level} detailing concrete design decisions, and the \emph{run time level}. Detail descriptions for each level are given in Table~\ref{05:tab_VEDLIoTLevels}.\par 

The \emph{run time level of abstraction} is of special interest because some requirements of an AI system might not be exhaustively testable before deployment. 
Russel describes in his book \emph{Human Compatible: AI and the Problem of Control} the example of an AI algorithm commonly found in social media that maximises \emph{click-through}, i.e., ``the probability that the user clicks on the presented items"~\citep{Russel2020}. 
Russel highlights that such an algorithm not necessarily ``presents items that the user likes to click on", but instead could (inadvertently) change the user's behaviour in a manner to make him or her more predictable in his preferences e.g., by favouring extreme political views~\citep{Russel2020}.
By constantly monitoring the decisions of the AI algorithm, such deviations from the intended behaviour can be detected and mitigated, e.g., through retraining or by ``pulling the plug". 
Most AI systems are not ``adaptive". 
They are trained and tested with a dataset representing the desired context in which the AI system is intended to operate in under the assumption of stationarity in the probability distribution of the data. 
In reality, the assumption of stationarity of the probability distributions does not hold in most case, for example when the context, in which the AI operates in, can change over time. 
Concepts like \emph{continual learning} allow the AI to handle drifts in data distributions~\citep{Lesort2021}. 
However, continual learning requires run time monitoring concepts to detect deviations from the currently learned context, and automatic data collection (and labelling) for autonomous retraining of the AI model. 

\begin{figure*}[t]
    \centering
    \includegraphics[width=\textwidth]
    {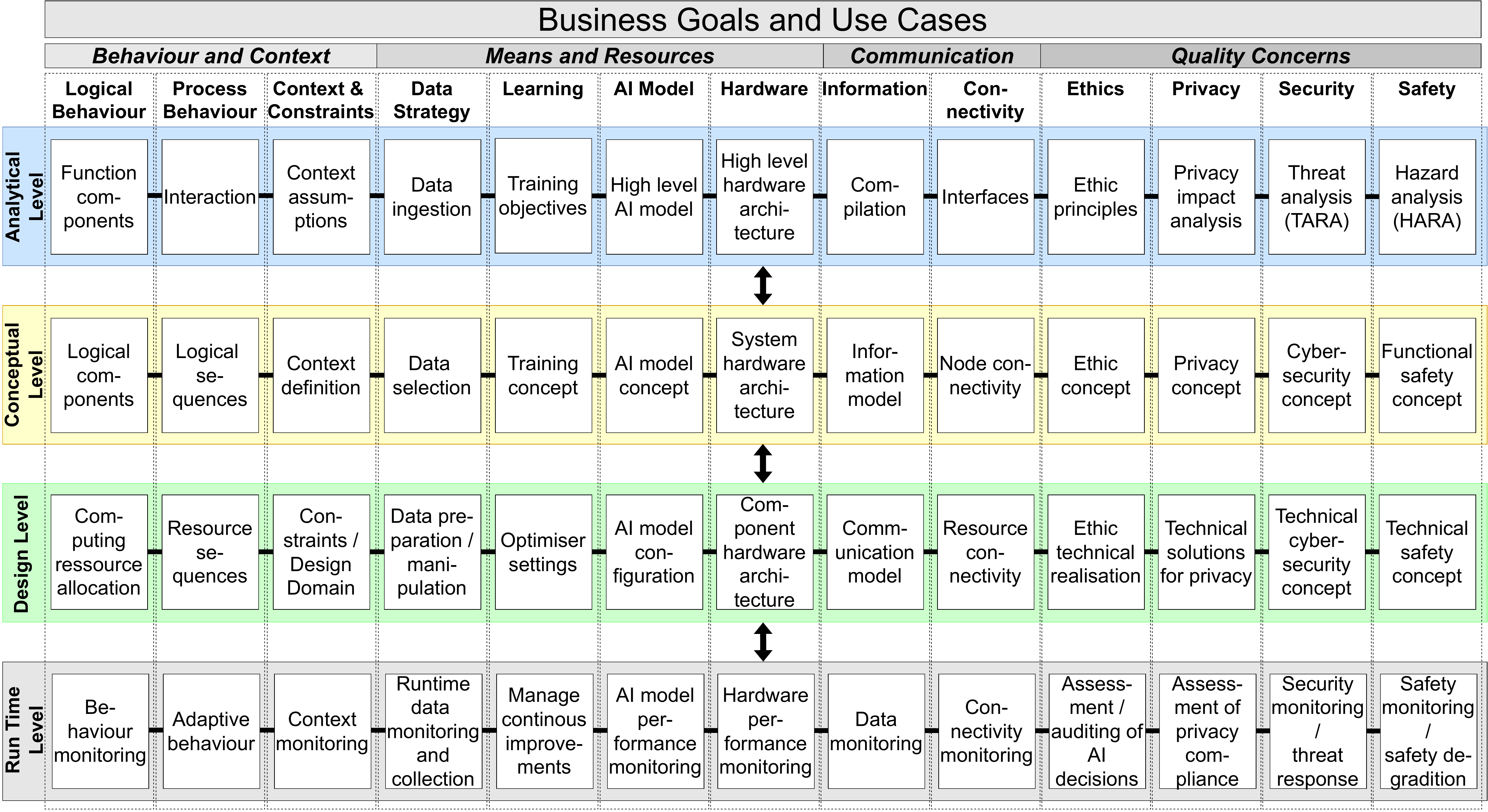}
    \caption{An architecture framework for VEDLIoT categorising views in different clusters of concern on different levels of abstraction. 
    }
    \label{fig:04_architecture_framework}
\end{figure*}

\subsection{Architectural views}
Finally, the workgroup populated the matrix with architectural views for the different clusters of concern at different levels of abstraction. The final matrix of architectural views for VEDLIoT is given in Figure~\ref{fig:04_architecture_framework}. 

Systems using the VEDLIoT toolchain will contain a significant amount of ``traditional" system components around the AI components in order to facilitate the desired behaviour. In this paper, we will not detail architectural viewpoints corresponding to these concerns since they are well covered in literature, e.g., in~\citet{IEEE2413} or the extensive viewpoint catalogue in~\citet[Part III]{Rozanski2012}.
However, other architectural viewpoints that aim to facilitate the design of AI components for the system are novel and will be the focus of this section.
Table~\ref{05:tab_ArchiViews} in \ref{app:descriptionarchiviews} provides a list of viewpoints, which govern architectural views in the architecture framework for VEDLIoT, that we assume to be relevant specifically towards the AI components of the system. Note that, we do not specify pre-defined models or diagrams suitable for the views because it is not the intention of a compositional architectural framework to provide a catalogue of architecture viewpoints and their models. Instead, the proposed framework aims at providing a method for constructing a coherent architectural framework with architectural views defined through the concerns relevant to fulfil all necessary requirements of the use case. It is the choice of the system architects, developers, and other stakeholders which viewpoints and models are suitable to create the views. Section~\ref{sec:06_demonstration} provides some example of diagrams for the AI specific views of one particular use case, but other use cases might want to use different viewpoints with different models.


In the proposed architectural framework, quality concerns are explicitly represented by architectural views on each level of abstraction. From a traditional system architecture point of view, one might question how, for example, a hazard analysis in form of a \emph{Hazard and Risk Assessment (HARA)}, can be constituted an architectural view. 
\citet{Tekinerdogan2011} proposed an approach to explicitly include \emph{quality architectural views} into architectural frameworks to overcome challenges when matching quality concerns with architectural elements representing functional aspects of the system.

The main challenge is that one cannot meaningfully analyse different quality aspects of the system without some prior knowledge of the system's architecture. On the other hand, improving quality aspects often entail changes in the system's functional architecture. \par
To give an example, we revisit the need of including the hazard analysis in the architectural framework: A work product that serves as input to a hazard analysis in accordance with ISO 26262 is the \emph{Item Definition}. The item definition contains a boundary diagram that depicts the elements of a system which are ``in scope" of the safety relevant system. This includes all functional components. Defining which elements are included in the boundary diagram is an architectural decision, and requires information provided through other, mostly functional, views of the system (i.e. there exist morphisms from functional views towards the hazard analysis view). The hazard analysis will return a set of safety goals that the ``item" has to achieve. How the safety goals are achieved is conceptualised in the \emph{Functional Safety Concept}. A common solution is to distribute the safety risk through safety decomposition, which often leads to the introduction of redundancies in the system. Again, introducing redundancies is a clear architectural decision, which has direct consequences e.g., for the system hardware architecture (i.e., we create a morphism from the Functional Safety Concept towards the system hardware architecture view). 
\par
Figure~\ref{fig:04_architecture_framework} presents the compositional architectural framework for VEDLIoT that aims at mitigating the identified concerns for distributed AI systems presented in Section~\ref{sec:03_challenges}.

\begin{figure}[!htb]
    \centering
    \includegraphics[width=0.8\linewidth]{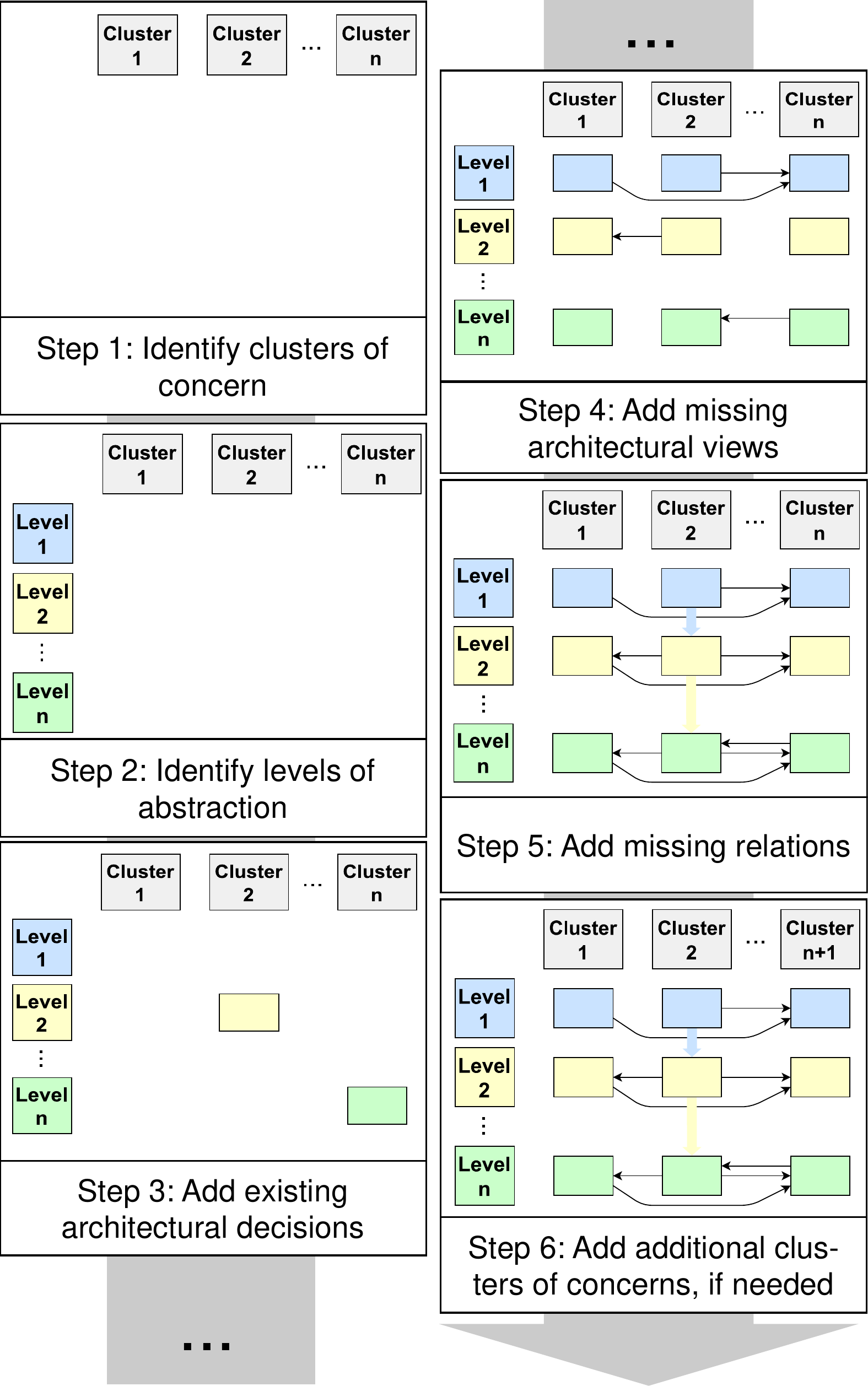}
    \caption{Steps for defining a compositional architectural framework}
    \label{fig:04_HowtoSteps}
\end{figure}

\subsection{Defining a compositional architectural framework}
\label{subsec:definecomp}
Based on the experience of applying a compositional architectural framework to VEDLIoT, the following guidelines can be provided:

{\small \begin{itemize}[leftmargin=4mm]
    \item \emph{Step 1:} Clusters of concern are identified based on the use case and business goals. Initially, larger groups of concerns (such as functionality, hardware, communication, quality) can be defined, which are then refined into atomic clusters of concern (Definition~\ref{prop0}).
    \item \emph{Step 2:} Levels of abstractions are identified. The number of required levels, and the level of detail on each of these levels depend on the size and complexity of system-of-interest and the development settings of the company. Three to four different levels of abstraction seem a good default (Definition~\ref{prop1}).
    \item \emph{Step 3:} Known architectural decisions are entered into the matrix. Most development projects do not start from scratch, but instead have to reuse or integrate into existing architectures. Prior knowledge, such as an existing component architecture, can be entered into the appropriate clusters of concern and levels of abstraction in the architecture matrix.
    \item \emph{Step 4:} Architectural views are added. Relations (morphisms) are created between the architectural views at each level of abstractions (Definition~\ref{prop1}) such that no inconsistencies occur when looking at the system-of-interest from different architectural views (Definition~\ref{prop2}).
    \item \emph{Step 5:} All relations between architectural views must be mapped onto corresponding views of the next lower level of abstraction (Definition~\ref{prop3}). If a relation between two architectural views on a higher level of abstraction does not have a correspondence on the next lower level of abstraction, the relation might be unnecessary and can be removed, or a corresponding relation needs to be created.
    \item \emph{Step 6:} During the system development, additional clusters of concern might be discovered and they are iteratively added. 
\end{itemize}}

\vspace{0.5em} \par \noindent All steps are illustrated in Figure~\ref{fig:04_HowtoSteps}. 

\section{Demonstration of the proposed framework on a use case} \label{sec:06_demonstration}

This section presents the results from applying the proposed compositional architectural framework for VEDLIoT on the development of an automatic emergency braking system, one of the use cases of VEDLIoT. The aim of this demonstration is to answer Research Question RQ3, which asks how a compositional framework can be applied in a realistic context.\par
Systems that mitigate frontal collisions are considered standard equipment in road vehicles today and standards exists that help design and test these systems \citep{ISO22839}. 
However, today's system yet do not exploit all the opportunities given by advanced AI.
For example today's systems are very limited in their capability of differentiating what specific object is detected as obstacles, in which context the vehicle is operating in, and how the hardware in the entire vehicle, and beyond on the edge and in the cloud, can be used more efficiently for processing-intensive tasks.\par
The use case serves as a detailed scenario to demonstrate the utility of the proposed artefact, i.e., the compositional architectural framework. Specifically, the aim is to demonstrate how the different challenges outlined in Table~\ref{tab:03_challenges} can be solved in a real-world scenario with the proposed compositional approach to an architectural framework.
We demonstrate how the logical behaviour, the distribution over different processing notes, and the hardware components can be planned concurrently using the compositional architectural framework approach.
The demonstrator also shows how deep learning models can be designed concerted to other design decisions, such as context definition, data strategy, and learning concept.
Lastly, we demonstrate how a quality concern such as safety influences other architectural views through morphisms between views.
The presented demonstrator entails only a small part of the use case development in VEDLIoT and does not cover the entire system development, because this would be outside the scope of this study. For a more detailed description of the use case development in VEDLIoT, the reader can refer to the VEDLIoT project and its deliverables, such as \citet{Berge2021}. 

\subsection{Evolution of logical components and hardware architecture}\label{subsec:UseCase2_BPC}
\begin{figure*}[!t]
    \centering
        \centering
        \noindent
        \includegraphics[width=\hsize]{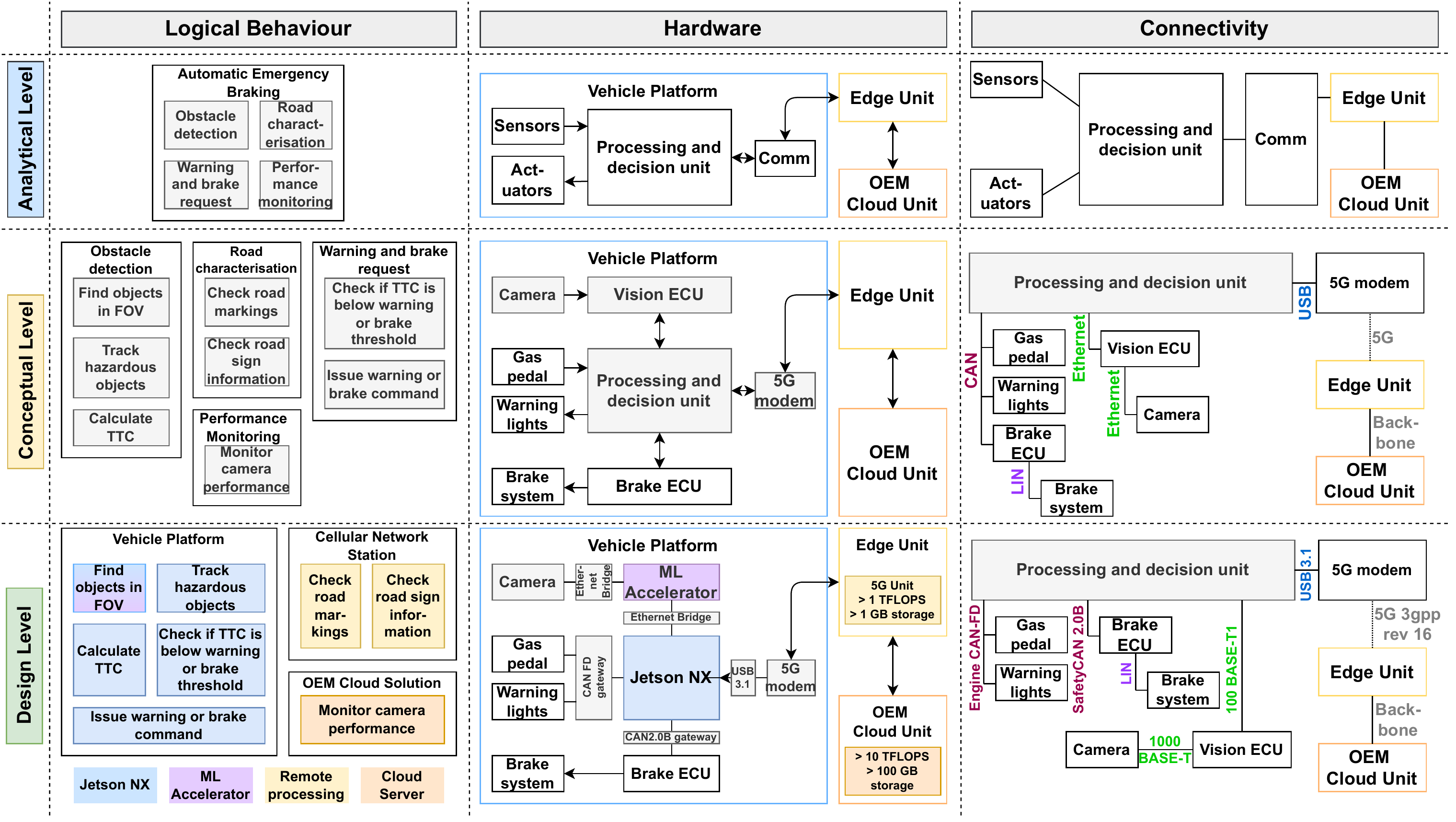}
        \caption{Co-Design of logical components, hardware design, and connectivity.}
        \label{fig:04_UC2_BP}
\end{figure*}
Figure~\ref{fig:04_UC2_BP} illustrates a simple example of the idea of co-design using the compositional architectural framework concept. 
The figure is an extract of the overall system architecture and depicts the logical behaviour design, the hardware architecture, and connectivity. 
On the highest level of abstraction, the function components view contains four main logical components: Obstacle detection, road characterisation, warning and brake request, and performance monitoring. 
Also, a high level view of the hardware architecture and connectivity is provided. 
On the next level of abstraction, the conceptual level, the function components are refined into logical components. 
Concurrently, the hardware architecture is refined into more details and first correspondences / morphisms are created: E.g., the need to use a visual camera in the system hardware architecture will cause a morphism towards the ``find objects in FOV" component, because the required algorithms for finding objects can be technology depending. A concept of connectivity between the hardware components is drafted; it outlines the desired network interfaces and connections. There are now morphisms between the system hardware architecture and the connectivity concept because changes of hardware components can require changes in the node connectivity.
On the design level of abstraction, the component hardware architecture provides detailed information on the hardware components. 
This is required, because the computing resource allocation view assigns the logical components to the different available hardware components. 
Several morphisms exist now between the computing resource allocation view and the component hardware architecture. 
A consistent system solution can only be obtained if the logical components are mapped to the hardware components represented in the component hardware architecture view. 
If logical components were mapped to hardware component that do not exist (let's say an additional ARM processor core), the morphism between the component hardware architecture and the computing resource allocation would be broken, and a consistent system solution could not be obtained. 
Furthermore, knowing the component allocation to hardware allows for specifying bandwidths and latency needs in the design level connectivity view.
This example showed how the identified challenges {\#6: \em Management of dependencies} and {\#9: \em Support of different levels of system design}, depicted in Table~\ref{tab:03_challenges}, are solved by the application of the compositional architectural framework.

\subsection{Correspondences between context, data strategy, learning, AI model, and hardware.}\label{subsec:CLAI}
\begin{figure*}[!t]
    \centering
        \centering
        \noindent
        \includegraphics[width=\hsize]{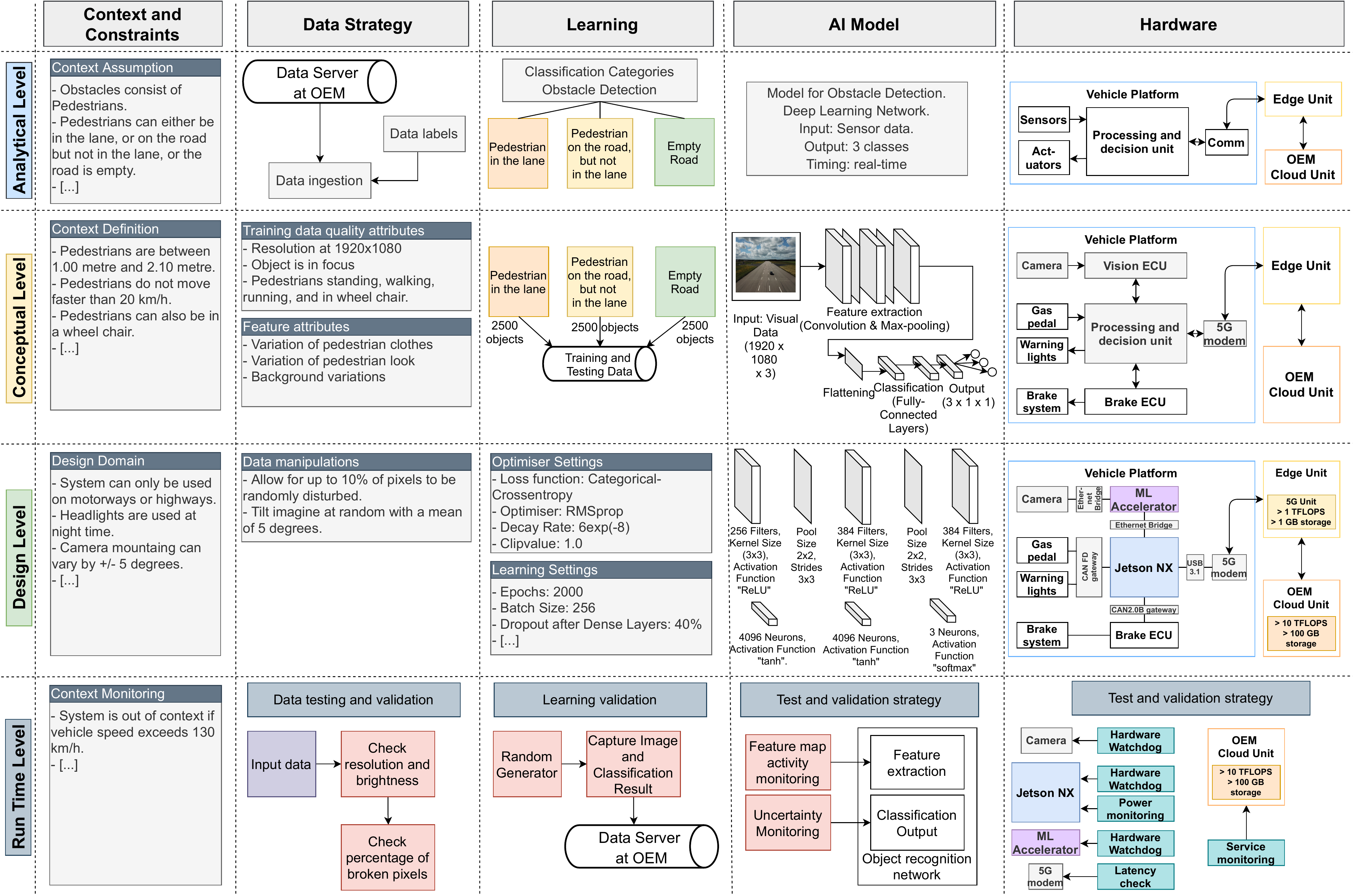}
        \caption{Co-Design of context and constraints, data strategy, learning concept, AI model, and hardware.} 
        \label{fig:04_UC2_CLAI}
\end{figure*}

The second example, provided in Figure \ref{fig:04_UC2_CLAI}, illustrates the parallel evolution of the context and constraints of the system, the data strategy, the learning concept, and the AI model. 
This example demonstrates how the development of an AI component can be seen as a hierarchical process which needs to stay in synchronisation with the remaining concerns of the system development. \par
On the highest level of abstraction, the context assumptions take direct influence on the required learning objectives (i.e., there exists a morphism from the assumption (``Pedestrians can either be in lane, or on the road but not in the lane, or the road is empty" to the learning object of classification of objects into three classes).\par 
On the conceptual level, the context definition clarifies the earlier context assumptions. This provides input to the data selection view in which feature attributes are specified for the data.
By establishing morphisms from the system hardware architecture view, the context definition can also take into account limitation of the hardware, e.g., the mounting position of the camera influences the minimum height of a human (e.g., 1.00 metre) to be detected. 
Now, the AI model can be conceptualised because input data, required output data and objective of the AI model are known. \par
On the design level, the design domain view provides clear constraints for the system's operability. Furthermore, the learning procedure's and AI model's configuration are set. This also includes a view on necessary data preparations or data manipulations. For example, after some time in operation the camera might suffer from random pixel failures. These pixel failures can be simulated by randomly disturbing pixels in the training data. The result is an  AI model that is more robust against random pixel failures.
Finally, the run time level provides views that explain which monitoring concepts and run time reconfiguration might be required during operation of the system.
We can take the AI model run time view as an example: Two monitors can check the feature map activity in the feature extraction section of the deep neural network, and uncertainty monitoring can be applied to the classification output of the network.\par
In summary, this example showed how the challenges (see Table~\ref{tab:03_challenges}) {\#1: \em Additional views for AI modelling}, \#2: {\em Consideration of data requirements}, \#3: {\em Context definition}, \#4: {\em Description of the AI's learning environment}, and \#8: {\em Run time monitoring} are mitigated by the architectural framework.

\subsection{Example of correspondences between a quality concern and the remaining architecture concerns}\label{subsec:SafetyExample}
\begin{figure}[t]
\begin{subfigure}[t]{\linewidth}
    \centering
    \includegraphics[width=\linewidth]{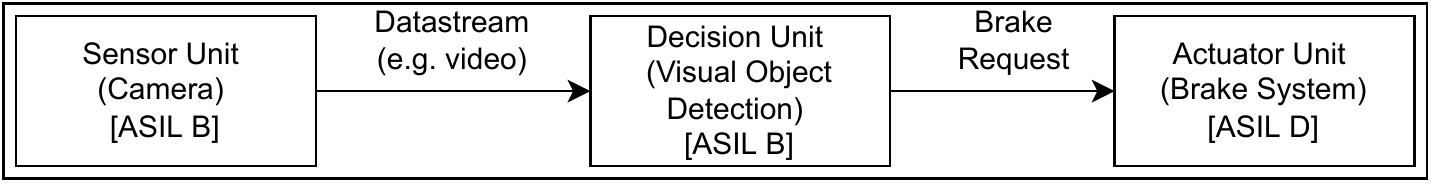}
    \caption{System architecture before safety decomposition.}
    \label{fig:04_SafetyA}
\end{subfigure}
\begin{subfigure}[t]{\linewidth}
    \centering
    \includegraphics[width=\linewidth]{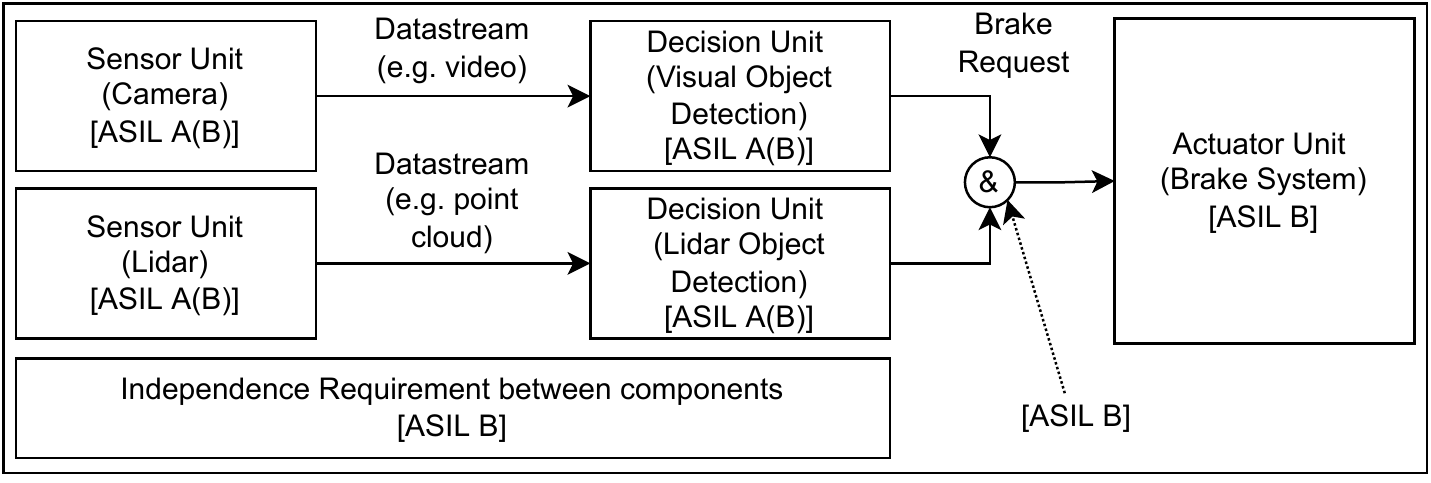}
    \caption{System architecture after safety decomposition.}
    \label{fig:04_SafetyB}
\end{subfigure}
\caption{Safety decomposition in the system architecture for an automatic emergency brake system.}
\end{figure}

This example demonstrates how a quality concern influences via morphisms other architectural views. This is an example on how quality aspects (even novel aspects such as explainability) can be explicitly handled in the architectural framework (Challenge~\#7 in Table~\ref{tab:03_challenges}).
A common quality concern for automotive systems is safety.
Assume the system shall trigger the brakes whenever a person is detected in the lane in front of the vehicle.
Assume further that, through the HARA, the following safety goal has been identified: \emph{``The system shall not trigger the emergency brake unintentionally (ASIL\footnote{Automotive Safety Integrity Level} B)"}. 
An extract of the high-level system architecture is illustrated in Figure~ \ref{fig:04_SafetyA}.
While the brakes are designed to a high safety integrity level, the camera and object detection algorithm might not be able to achieve the required ASIL B. 
Therefore, a safety decomposition in accordance with ISO~26262 results in redundancy in the sensing system, and a lower ASIL on each component: 
In the functional safety concept, an additional lidar sensor, together with a second object detection algorithm specifically designed for detecting objects in lidar point clouds, allow for the reduction of the required safety integrity level of all redundant components to ASIL~A(B). 
The additional sensing system must be independent from the first object detection algorithm.
The final high level system architecture after safety decomposition is illustrated in Figure \ref{fig:04_SafetyB}. 

By introducing the safety decomposition in the functional safety concept, correspondences (morphisms) to the system hardware architecture view and the logical components view were established. 
On the next level of abstraction, the technical safety concept establishes a view on the overall system's architecture that allows the fulfilment of the functional safety concept. 
For example, the technical safety concept provides a view on the system architecture that requires the logical component ``Visual object detection" to be deployed on safety certified hardware components, which creates a correspondence to the computing resource allocation view; or that the object detection algorithm only works at daylight, which creates a correspondence to the constraints / design domain view. \par
A major advantage of the compositional architectural framework is the ability to trace morphisms (i.e., links or correspondences) between the different architectural views. Assume, that after years of service the system's hardware shall be upgraded and it is decided to replace the two independent processing units with a more powerful single unit. Then, the morphism between the system hardware architecture and the functional safety concept reminds the system designer of the safety concern that triggered the original design decision of having separated and independent processing units for the two independent object detection algorithms, even years after the original development.

\subsection{Feedback from industry partners in VEDLIoT}
\label{subsec:feedback}
In this article, we presented only one of four use cases to which the compositional architectural framework approach was applied to. The other use cases were a fault detection system for high voltage switching, a system for electric motor condition classification, and a smart mirror as part of a smart home setup. Because it is beyond the scope of this article to detail all demonstrator, we used semi-structured interviews to collect feedback on the usability, the consistency of the compositional architecture framework for VEDLIoT, as well as feedback on possible improvements, how well the approach fits with their use case, and comparisons to current system architecture frameworks applied in the company. We interviewed four use case developers, one from each use case. The interviews allowed us to gain opinions and feedback from participants who had experienced the compositional architecture framework and to explore yet un-identified issues with the subject \citep{Hancock2006}. The interviews contained three sets of questions: I) Questions about the interviewee's role and experience with architecting complex systems; II) A set of feedback questions on the usefulness and applicability of the compositional architecture framework; III) The role of run time monitoring in the VEDLIoT systems and their architectures. The interview guide with all questions is available in the replication package of this article\footnote{\url{https://doi.org/10.7910/DVN/VXFFFU}}. Due to data privacy reasons, contact information to the use case owners is not included in this article. But general contact information to the companies can be found on the website of the VEDLIoT project\footnote{\url{www.vedliot.eu}}. \par
\paragraph{Background of the interviewees} The four interviewees have all a background in system architecting with different years of experience. One interviewee was a PhD candidate in charge of developing a prototype system, two interviewees work as project leader and system developer for a multinational company ($>$10.000 employees) with 4 and 10 years of experience, and one interviewee is a research specialist for system design in an automotive supply company ($>$1.000 employees) with more than 20 years of experience in system development.
\paragraph{Challenges solved through the architectural framework} 
We presented the challenges discussed in Section~\ref{sec:03_challenges} to the interviewees and asked which of the challenges they encountered and in what degree the architectural framework helped mitigating them.
All interviewees mentioned that the explicit treatment of quality aspects as part of the architectural framework was helpful. It eased finding trade-offs between competing quality aspects, and it allowed for better cooperation between different teams dealing with different quality aspects of the system.\par
Having explicit cluster of concerns for data strategy and learning helped all partners in defining data sets and training configurations. One partner mentioned, that, in contrast to previous in-house processes they used, the architectural framework established a much better connection between the context description, the hardware architecture, the AI model and the data sets used for training. Establishing connections early between the context and the data strategy reduced the need for data creation in a laboratory, which reduced the costs of the project.\par
We asked in what degree the guidelines of the architectural framework helped them in finding the right architecture for their system, compared to other approaches or processes they usually apply in their respective company.
A common answer was that, compared to previous architectural framework they used, the guidelines helped in keeping an overall structure and overview of the system development. The differentiation into different level of abstractions prevented that design decisions are taken too early, which could have created boundaries in the later system development. Also, one interviewee highlighted that the proposed framework fit better into an agile development environment, because the guidelines can help different teams in keeping a consistent architecture.\par
Furthermore, according to two interviewees, the guidelines established an easier traceability of design decisions, which helped in documenting causes for architectural decisions and in fulfilling documentation requirements for certification (e.g., for fulfilling safety standards, or ethical aspects of AI as governed by new EU regulations) of the products.\par
All use case owners agreed that the explicit treatment of run time behaviour in the architectural framework is helpful. One use case owner used the views in the run time level to design explicit feedback mechanisms allowing the user to report on the run time experience of the system. Two use cases created run time views that ensured that legal regulations are met and controlled at run time.  
\paragraph{Missing aspects or negative experiences} We asked if any information or aspects were missing in the architectural framework. One interviewee mentioned that it was difficult to decide on the importance of the cluster of concerns and where to start the system design process. The interviewee suggested to highlight those clusters of concern that are most likely compulsory for each system development (e.g., logical behaviour, context and constraints, and hardware) and use them as starting point for the system development. Another use case owners said he wished for more software architecture concerns, such as views on the operation system and middleware. Furthermore, the same use case owner said that, although the guidelines helped significantly in establishing traceability of design decision for certification of e.g., safety aspects, an explicit treatment of certification and standardisation aspects in the framework would be appreciated. Lastly, one use case owner mentioned that a cluster of concern could be established on human-machine-interfaces, containing views on the interaction with the users or operators of the system. All use case owners agree that some form of tool support for the architecture framework would be highly appreciated.

\section{Discussing the relation to requirement engineering} \label{sec:07_discussion}
The three examples of applying the architectural framework given in Section~\ref{sec:06_demonstration} show an interdependence between the systems engineering and requirements engineering. For example, the functions selected for deployment on the Jetson NX (highlighted in blue on the Design Level in Figure~\ref{fig:04_UC2_BP}) set performance requirements on the processing unit (Jetson NX).
Indeed, based on the twin peaks model, \citet{Nuseibeh2001}, emphasise in their hierarchical requirements reference model the importance of the interrelation between system design and requirements. We realised that the compositional architectural framework provides a refinement structure, which complements and supports requirement engineering.

\subsection{Traceability of design decisions}
\begin{figure}[!t]
    \centering
        \centering
        \noindent
        \includegraphics[width=\linewidth]{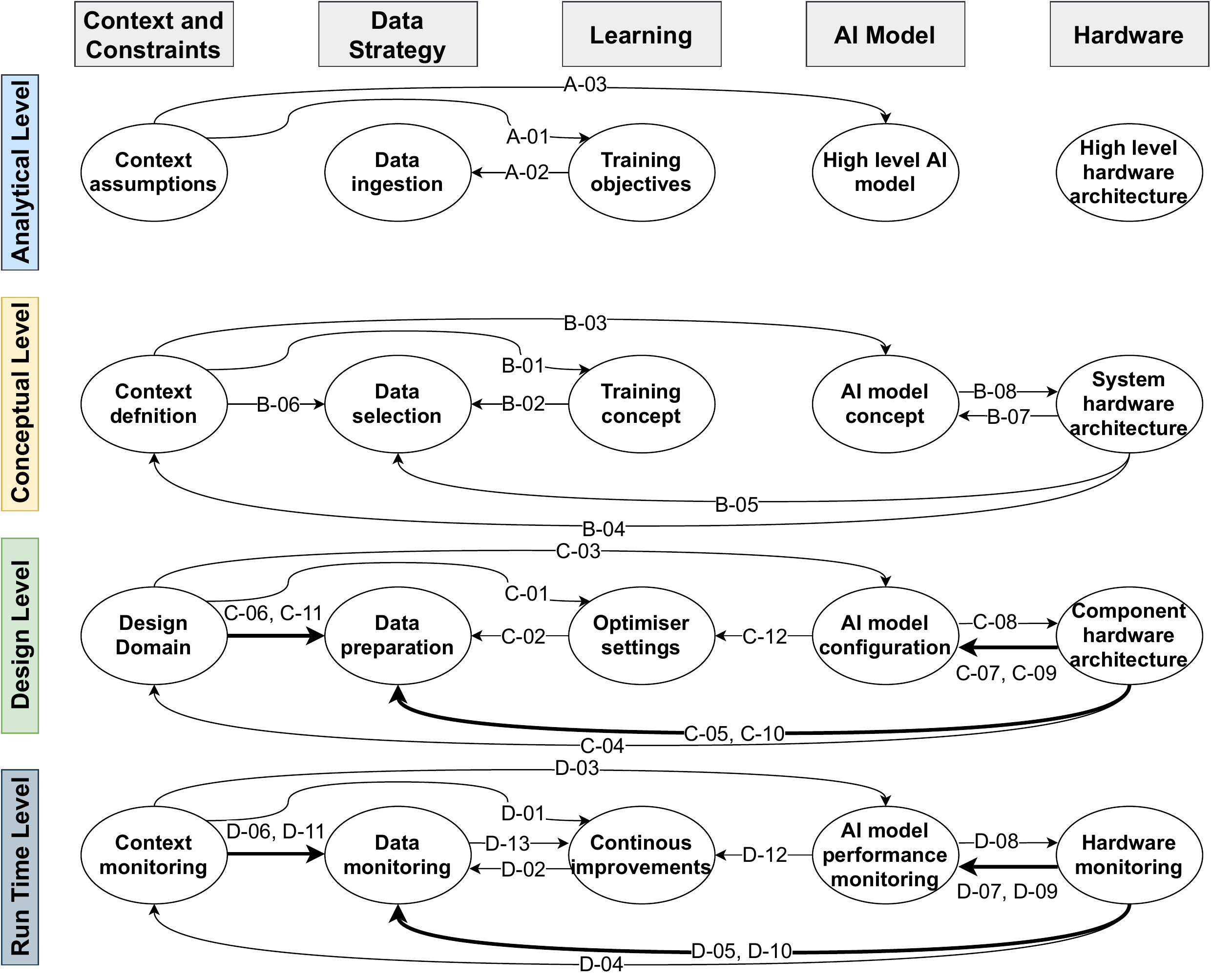}
        \caption{Directed graph illustrating the evolution of relations between architectural views in the compositional architectural framework. A table of all relations for this example is available in~\ref{app:relations}} 
        \label{fig:04_UC2_CLAI_corr}
\end{figure}

Based on the feedback from the use case owners on applying the compositional architectural framework in VEDLIoT, we learnt that the framework helped in establishing traceability in design decisions through morphisms between the architectural views. Figure~\ref{fig:04_UC2_CLAI_corr} shows how the relations between the architectural views evolved for the example discussed in Section~\ref{subsec:CLAI}. It illustrates that design decisions made in one architectural view can cause requirements on elements in another architectural view. Through formalising the correspondences in the architectural framework, traceability of the design decisions can be established. Therefore, it extends existing architectural frameworks, such as Viewpoints + Perspectives \citep{Rozanski2012}, by providing a mathematical foundation to establish traceability of design decisions in the architectures through the rules outlined in Section~\ref{sec:04_framework}.

\subsection{Support of middle-out design}
Traditionally, requirement engineering would be organised in a top-down fashion. 
However, the architectural framework supports middle-out systems engineering, which is a widely common practise, combining traditional top-down systems design with integration of designated lower level hardware, software, AI models, or other components~\citep{Davis2019}. The need for middle-out design has also been identified as Challenge~\#10 in Table~\ref{tab:03_challenges}.
Knowledge can become available on all levels of the architectural framework at any time.
For example, a new desired function shall utilise an existing AI model and run on a predefined hardware platform. 
Thus, the aim of requirement engineering is to ensure completeness of missing views on a conceptual or analytical level based on existing design level views.\par
Another example is that of user stories: User stories, can mix problem and solution descriptions. 
For example, \emph{``as a driver, I want the vehicle to brake automatically so that I do not hit an obstacle on the road"}, transmits a problem to be solved. 
However, \emph{``as product owner, I want the automatic emergency braking function to execute on an existing hardware platform."} transmits a solution description.
Because the compositional architectural framework supports middle-out development, the solution descriptions would influence the design level, while the problem descriptions affect the analytical level of the architecture description.

\subsection{Defining the scope of the system}
The concrete design domain of a solution is found by iterating between the problem and solution space. 
The analytical level of the architectural framework accumulates information about the problem space. 
The further the design proceeds, the more information from a solution space perspective is added to the system architecture and its design domain. 
This became evident in the example shown in Section~\ref{subsec:UseCase2_BPC}: Concrete hardware solutions which are part of the solution space are defined only on the design level of the architecture (e.g., the selection of a Jetson NX platform in Figure~\ref{fig:04_UC2_BP}). 
The conceptual level is mostly, and the analytical level is completely hardware agnostic and therefore they represent information from the problem space.
Finally, in the design phase, it mostly will be the solution space that provides additional information. 
As an example, assume a certain hardware component can only operate correctly in a temperature range of -30 degrees to 60 degrees. 
The component hardware thus limits the design domain in regards of allowed temperature range. This would also create a correspondence to the \emph{``Constraints / Design Domain"} view in the \emph{``Context \& Design"} cluster of concern, and thus would make this constrain, and its relation to the hardware component, explicit in the architecture (Need of context and design domain descriptions, Challenge~\#3 in Table~\ref{tab:03_challenges}).


\subsection{Ensuring the desired behaviour of an AI system} 
The behaviour of an AI system depends on the available data, and the context. 
\citet{Rao2021} therefore propose to reference the context and data definitions together with the use cases. 
We saw in the use case demonstration in Section~\ref{subsec:CLAI} that all elements of the information model proposed by Rao et al. can be represented by the compositional architectural framework:
Context, and Context Elements are described in the \emph{Context and Constraints} cluster of concern; data requirements and data sources are described through \emph{Information} cluster of concern, and the {Learning} cluster of concern; and quality attributes are represented through the group of \emph{Quality} clusters of concern. 
Explicit knowledge of the necessary data quality and context of operations is a prerequisite for defining run time monitors, which was depicted as Challenge~\#8 in Table~\ref{tab:03_challenges}.

\subsection{Defining quality goals}
Quality goals can be distinguished into quality attributes (such as performance requirements, or specific quality requirements) and constraints \citep{Glinz2007}. 
A first step to determine which quality attributes need to be represented in the architectural framework explicitly in the form of clusters of concern, a quality grid analysis can be formed \citep{Lauesen2002}.
The quality grid is essentially a table with relevant quality attributes in the first column and then an indication on whether this attribute is more or less important than in comparable products. 
Quality attributes that rank high in this analysis can then be made a cluster of concern in the compositional architectural framework.
By representing a quality attribute as explicit quality concern in the architectural framework, the fulfilment of that quality attribute becomes part of the overall system design process (i.e., safety-by-design, security-by-design, etc). This eases the inclusion of novel quality concerns, such as explainability and the integration of additional stakeholders, such as social scientists (ethical requirements) or public authorities (privacy requirements) in the system development (Challenges~\#5 and \#7 in Table~\ref{tab:03_challenges}). Their concerns are explicitly reflected as clusters of concern in the architecture of the system.

\subsubsection*{Open targets for quality requirements} 
An additional advantage of including quality attributes as concerns with different levels of abstraction is the ability to support open metrics, or open targets. 
According to \citet{Lauesen2002}, it is good practice to defer the selection of metrics and its target values for quality attributes to a late stage in the systems development. 
This strategy avoids a false sense of accuracy and eventually over-designed solutions. 
The quality concerns can contain target values as late as in the design level of the compositional architectural framework. 
At the highest level, the analytical level, it is only important to understand what quality attributes are relevant for the general system design.\par


\section{Conclusion}
\subsection*{Answer to RQ1: Which challenges are relevant when defining system architectures for AI systems?}
The article outlined challenges and concerns when developing systems that can include AI components. Together with industry partners, these concerns were collected for distributed systems with deep learning components. Specifically, the empirically identified challenges include concerns related to data quality aspects, AI modelling, or setting up a correct learning environment for the AI. Additionally, existing quality aspects such as security and safety have to be combined with novel aspects such as ethical considerations or explainability. New stakeholders enter the stage of systems engineering, such as data engineers which bring their own views (on the system) and language. This creates a huge variety of different architectural views, for which existing standards for architectural frameworks are not sufficient. Especially identifying and mapping of dependencies between the architectural views has been identified as a critical challenge.

\subsection*{Answer to RQ2: What guidance can compositional thinking provide to overcome these challenges for the design and management of architectures for AI systems?}
For the purpose of defining and managing architectural views, a theoretical approach to a compositional architectural framework based on ideas from category theory was proposed. Four suggested propositions can be translated into rules that provide guidance for the creation and management of a compositional architectural framework, see the \emph{Rules for compositional architecture} box on this page.

\begin{tcolorbox}[floatplacement=t,float,title=Rules for compositional architectures]
\footnotesize
\emph{A cluster of concern is partially ordered set of architectural views which represent a specific concern of the system at different levels of abstraction.} \par
\begin{adjustwidth}{2mm}{}
\textbf{Rule 1: Clusters of concern shall contain architectural views with different levels of details of a certain aspect of the system-of-interest.} \par
\end{adjustwidth}

\vspace{0.5em} \noindent \emph{The set of architectural views with equivalent level of details about the system-of-interest constitutes a category called level of abstraction. Architectural views on a level of abstraction are related to each other through morphisms. Morphisms exist only between architectural views on the same level of abstraction.} \par
\begin{adjustwidth}{2mm}{}
\textbf{Rule 2: Architectural views shall be sorted into levels of abstraction. Views on the same level of abstraction shall have an equivalent level of details about the system-of-interest. Correspondences shall exist only between architectural views on the same level of abstraction.}\par
\end{adjustwidth}

\vspace{0.5em} \noindent \emph{If the product over all architectural views on a level of abstraction is valid, the architectural views consistently describe the system-of-interest.} \par
\begin{adjustwidth}{2mm}{}
\textbf{Rule 3: By using correspondence rules, it shall be possible to arrive at different architectural views of the system-of-interest without encountering inconsistencies.}\par
\end{adjustwidth}

\vspace{0.5em} \noindent \emph{Functors map all views of one level of abstraction to corresponding views of the next lower level of abstraction, and all relations between views to corresponding relations of the next lower level of abstraction.} \par
\begin{adjustwidth}{2mm}{}
\textbf{Rule 4: Architectural views, and relations between them, on a higher level of abstraction shall be mapped onto corresponding views and relations on the next lower level of abstraction.}
\end{adjustwidth}
\end{tcolorbox}

Unlike previous approaches to architectural frameworks for the IoT and AI systems, such as the IEEE 2413 standard \citep{IEEE2413}, the proposed idea for compositional architectural frameworks allow for system architectures that are scalable to the number of architectural views required for complex systems such as distributed AI systems. 
Furthermore, it provides guidance for managing dependencies between the architectural views, something that for example the 4+1 view model by \citet{Kruchten1995} is not considering. Rules based on mathematical considerations help in organising different views with different levels of details, keeping consistency between all views, and they help in ensuring traceability of design decisions. This extends existing architectural frameworks such as Viewpoints + Perspectives \citep{Rozanski2012}. We showed how the idea of a compositional architectural framework supports requirement engineering efforts for complex systems by explicitly supporting context descriptions of the system, middle-out system engineering, and any number of required quality goals. 

\subsection*{Answer to RQ3: How can a compositional framework be defined and applied in a realistic context?}
By following the proposed rules, a workgroup of 16~participants from academia and industry created a compositional architectural framework for the VEDLIoT project. The group started by identifying the necessary clusters of concern, levels of abstractions, and finally necessary architectural views for VEDLIoT. The created framework for VEDLIoT was demonstrated on a use case from the automotive industry. The demonstration showed that the framework can handle different architectural views for concerns such as logical behaviour, hardware, context, data strategy, learning environment, AI model, and safety aspects. The compositional framework integrates and complements requirements activities in a favourable way, thus supporting middle-out design and decomposition of requirements for AI systems. The feedback from four use case owners who applied the architectural framework in their projects was favourable. They highlighted that, compared to previously used frameworks and processes, the compositional architecture framework helped their development in keeping a more structured and better overview of the architecture, in finding trade-offs easier between quality aspects, in establishing traceability of design decisions, and in supporting an agile development of the system architecture.

\subsection{Threats to validity}
While we assume that the theoretical idea behind the compositional architecture framework is valid due to the mathematical definitions of the underlying principles, the application of these principles to VEDLIoT can contain threats to validity. First of all, VEDLIoT concentrates on the application of deep learning in distributed systems, which only represents one of many possible applications of the compositional framework. Furthermore, the composition of the workgroup and focus groups involved in the creation and discussion of the architectural framework for VEDLIoT contains mostly people from both academia and industry dealing with advanced research projects. Out of 16 participants, four participants work as developers for systems with deep learning in industry. While all other participants have significant knowledge in system architecture design, system development, and requirement engineering, experience ``from the field" of deep learning development could only be provided by 25\% of the participants. This potentially can be a cause of mismatch in the case study. By conducting review interviews with additional use case owners, we received feedback on the utility of the framework by additional industry practitioners. The interviews were also an effort to reduce the confirmation bias by including participants not involved in the creation of the framework.

\begin{table*}[!t]
\centering
\caption{Additional suggestions for validation, including hypotheses and actions.}
\label{tab:08_validation}
\footnotesize
\begin{tabular}{cp{15cm}c}
\hline
\rowcolor[HTML]{C0C0C0} 
\textbf{ID} &
  \begin{tabular}[c]{@{}l@{}}\textbf{Description} \end{tabular}&
  \begin{tabular}[c]{@{}c@{}}\textbf{Relates to} \\ \textbf{section} \end{tabular}  \\ \hline
1 &
The construction of the proposed architecture framework followed a clearly defined research methodology described in Figure~\ref{fig:02_structure}. The framework is built on awareness on the state of the art through a survey of published literature and standards. Furthermore, we identified the state of practice by conducting workshops within the VEDLIoT project, because the aim of the case study was to develop an architecture framework suitable for VEDLIoT. Validating the awareness of the state of practice outside of VEDLIoT requires empirical data from practitioners external to the project. &
  \ref{subsec:03_Empiricalchallenges} \\ \hline \rowcolor[HTML]{EFEFEF}
 & \multicolumn{2}{l}{\begin{tabular}[c]{@{}p{17cm}@{}}\textbf{Hypothesis 1:} The state of practice and specifically the challenges identified within VEDLIoT can also be representative for other project.\end{tabular}} \\ 
\rowcolor[HTML]{EFEFEF}
\begin{tabular}[c]{@{}l@{}}$\rightarrow$\end{tabular}& \multicolumn{2}{l}{\begin{tabular}[c]{@{}p{17cm}@{}}\textbf{Action 1:} Validate the state of practice with practitioners external to the VEDLIoT project.\end{tabular}} \\ \hline
2 &
A mapping between the theory of compositional architecture frameworks and the identified challenges of constructing systems within the VEDLIoT project is derived from both the state of the art as defined through literature and standards, and empirical data collected within VEDLIoT. The identified challenges themselves, and the mapping towards the theory of compositional architecture frameworks need to be validated outside of the VEDLIoT project in order to establish better generalisability of the mapping. &
  \ref{subsec:03_Empiricalchallenges}, \ref{sec:solutionChallenges}  \\ \hline \rowcolor[HTML]{EFEFEF}
 & \multicolumn{2}{l}{\begin{tabular}[c]{@{}p{17cm}@{}}\textbf{Hypothesis 2:} The mappings between the challenges and the theoretical framework can also be applied to other projects.\end{tabular}} \\ 
\rowcolor[HTML]{EFEFEF}
\begin{tabular}[c]{@{}l@{}}$\rightarrow$\end{tabular}& \multicolumn{2}{l}{\begin{tabular}[c]{@{}p{17cm}@{}}\textbf{Action 2:} Validate the mapping between the theory of compositional architecture frameworks and the challenges through independent experts. A possible experiment setup could take the form of different surveys that validate the challenges listed in Table~\ref{tab:03_challenges} and elicits additional challenges. An additional survey can then check the mapping between the theory and the challenges. \end{tabular}} \\ \hline
3 &
 In order to validate the instantiation of the compositional architecture framework within the VEDLIoT project, the participants of the project are most suitable, because they are the ones knowing the needs of the project and, therefore, they can assess the applicability and usefulness. This can depend on prior knowledge and usage of architecture frameworks, and therefore different projects might evaluate the applicability and usefulness of the proposed approach differently. & 
  \ref{subsec:definecomp}, \ref{subsec:feedback}  \\ \hline \rowcolor[HTML]{EFEFEF}
   & \multicolumn{2}{l}{\begin{tabular}[c]{@{}p{17cm}@{}}\textbf{Hypothesis 3:} The applicability and usefulness of the proposed approach to apply compositional architecture frameworks in system design can also be shown in other projects.\end{tabular}} \\ 
\rowcolor[HTML]{EFEFEF}
\begin{tabular}[c]{@{}l@{}}$\rightarrow$\end{tabular}& \multicolumn{2}{l}{\begin{tabular}[c]{@{}p{17cm}@{}}\textbf{Action 3:}  Instantiate the compositional architecture framework to further use cases and validate the success of this instantiation with the practitioners responsible for the new use cases based on their prior usage of architecture framework.\end{tabular}} \\ \hline
4 &
For validating the generalisability of the guidelines on how the theoretical framework is instantiated, independent of the case company and VEDLIoT project, there is the need of involvement of practitioners outside the project. 
   &
  \ref{subsec:definecomp} \\\hline \rowcolor[HTML]{EFEFEF}
   & \multicolumn{2}{l}{\begin{tabular}[c]{@{}p{17cm}@{}}\textbf{Hypothesis 4:} The guidelines on how to build a compositional architecture framework can be transferable to other projects.\end{tabular}} \\ 
\rowcolor[HTML]{EFEFEF}
\begin{tabular}[c]{@{}l@{}}$\rightarrow$\end{tabular}& \multicolumn{2}{l}{\begin{tabular}[c]{@{}p{17cm}@{}}\textbf{Action 4-a:} Apply the guidelines towards building a compositional architecture framework to another case.  \end{tabular}} \\ \rowcolor[HTML]{EFEFEF}
\begin{tabular}[c]{@{}l@{}}$\rightarrow$\end{tabular}& \multicolumn{2}{l}{\begin{tabular}[c]{@{}p{17cm}@{}}\textbf{Action 4-b:} Scrutinise each guideline by external experts, both via questionnaires and interviews to collect quantitative and qualitative aspects that might lead to identify risks and benefits of each guideline.  \end{tabular}} \\ \hline
5 &
For evaluating the applicability of the compositional approach towards architecture frameworks in a realistic context, there is a need of experts of the domain of that context. In Section~\ref{sec:06_demonstration}, we demonstrated the applicability on the development of an automatic emergency braking system. We used experts from the automotive domain to apply and validate the approach. Arguing for the applicability of the approach in other contexts requires additional use cases.
   &
  \ref{sec:06_demonstration} \\ \hline \rowcolor[HTML]{EFEFEF}
 & \multicolumn{2}{l}{\begin{tabular}[c]{@{}p{17cm}@{}}\textbf{Hypothesis 5:} A compositional architecture framework built on the described theory can also be applied in other realistic contexts outside the automotive industry.\end{tabular}} \\ 
\rowcolor[HTML]{EFEFEF}
\begin{tabular}[c]{@{}l@{}}$\rightarrow$\end{tabular}& \multicolumn{2}{l}{\begin{tabular}[c]{@{}p{17cm}@{}}\textbf{Action 5:} Apply the proposed approach towards compositional architecture frameworks in additional realistic contexts outside of VEDLIoT. \end{tabular}} \\ \hline
\end{tabular}
\end{table*}

\subsection{Additional steps towards validation}
The theoretical model of a compositional architectural framework was tested in practice through a case study conducted 
within a company from the automotive industry in the context of distributed deep learning systems. Additional practitioners outside of the case company confirmed through interviews the usefulness of the compositional approach described through the proposed mathematical model towards defining system architectures. Further validation steps can help in increasing external validity of the findings presented in this article. Especially selection bias can be a concern because we only applied the proposed mathematical concept of compositional architectures to one concrete case in the context of automotive distributed deep learning. Table~\ref{tab:08_validation} details additional validation steps, including references to different sections of this article and proposed actions for validation.

\subsection{Further research}
Besides the additional steps towards validating the proposed theoretical contribution to compositional architecture frameworks, further research can be conducted in how to apply category theory to system and software architectures. Category theory is a very broad field of mathematics, which is why it is difficult to find a starting point to develop ideas from it for a given problem. 
However, looking at the concepts of profunctors and monoidal categories, there are many interesting concepts to be found in applied category theory that could solve problems we today face in requirement engineering and systems engineering. 
This paper gives a starting point by introducing categorifaction to an architectural framework. 
The article's main focus was on evaluating this starting idea by creating a compositional architectural framework for systems with deep learning components. 
The VEDLIoT project includes an open call with the aim to apply, among other tools, the VEDLIoT toolchain to new use cases. 
The open call can serve to further validate the idea of a compositional architectural framework and to collect best practices on how to work with and apply the VEDLIoT architectural framework to new use cases. 
Further research is proceeding in developing software tools that support the ideas of a compositional architectural framework. One such tool could be built upon the idea of using textual architectural description and a distributed version control system such as \emph{git}, see for example \citet{Knauss2018}.

\section*{Acknowledgements}
This project has received funding from the EU’s Horizon 2020 research and innovation program under grant agreement No 957197 (vedliot). This work was also supported by the Centre of EXcellence on Connected, Geo-Localized and Cybersecure Vehicles (EX-Emerge), funded by the Italian Government under CIPE resolution n. 70/2017 (Aug. 7, 2017). We are thankful to all interviewees and companies who supported us in this research.
\FloatBarrier




\bibliographystyle{elsarticle-harv} 
\bibliography{cas-refs}

\appendix
\begin{onecolumn}
\section{Data produced for this study}
This section contains a description of the data collected during each cycle of the design science study.
\subsection*{Problem identification (Section~\ref{sec:03_challenges})}
\paragraph{Literature review}
The survey of literature was not conducted as a systematic literature survey. Instead, already existing literature surveys on software engineering method for AI-based system \cite{Martinez2021} and IoT architectures \cite{Ray2018} served as starting point. Literature pointing towards challenges and limitation of current architectural approaches were used to identified further relevant prior and derivative work through the tool \emph{Connected Papers}\footnote{https://www.connectedpapers.com/}. \par
\paragraph{Workshop 1: Identification of challenges}
A first workshop with a focus group of 11~experts from the VEDLIoT project was conducted on 17th February 2021. The list of participants is given in Table~\ref{tab:03_participants}. The workshop lasted 2 hours. It was conducted through the online meeting tool \emph{Zoom} and used the tool \emph{Mentimeter} for interactive elements. A recording of the workshop can be produced upon request. Prior to the workshop, the participants received a list of stakeholders and challenges as they are mentioned in the IEEE~2413 standard. This list was discussed during the workshop, and additional challenges or stakeholders added if needed.
\paragraph{Workshop 2: Confirmation of challenges}
A second workshop with the same focus group from VEDLIoT was conducted on the 31st March 2021. The same participants as listed in Table~\ref{tab:03_participants} joined through the online meeting tool \emph{Zoom}. The aim of the second workshop was to validate the findings of the first workshop. The workshop lasted for 2 hours. A recording of the workshop is available.
\subsection*{Creation of a compositional architecture framework for VEDLIoT (Section~\ref{sec:05_application})}
\noindent Applying the theory of compositional architectural frameworks to VEDLIoT was conducted through bi-weekly meetings with altogether 16~experts from both academia and industry. The meeting occurred every second Thursday between 25th February to 17th June 2021 (8 occurrences). The meetings were conducted online through the meeting tool \emph{Zoom}. Each meeting was recorded. Again, interactive elements such as \emph{Mentimeters} were used to collected opinions from the group of experts. 
\subsection*{Review interviews: Experiences from practitioners (Section~\ref{sec:06_demonstration})}
\noindent After the first four use cases of VEDLIoT applied the architecture framework, we invited the use case owners (i.e., project managers) for interviews to collect feedback on the usefulness and problems when applying the framework. We interviewed one project responsible per use case (i.e., 4 interviewees). Two interviewees were interviewed on-site, two interviewees via a remote Zoom meeting. An interview guide was created beforehand and only available to the interviewer. The interviewer took notes, and recorded the meetings.

\section{Description of the VEDLIoT project}
\label{app:VEDLIoT}
The VEDLIoT project has run since the beginning of 2021 as a cooperation between six academic and five industrial partners from different industry sectors who commonly face challenges when developing AI systems. 
VEDLIoT aims at creating a toolchain for the development of distributed (IoT) systems that employ deep learning models. Use cases from industry partners serve as demonstrators of the new VEDLIoT toolchain. 
Modern system development faces two major challenges: First, more and more traditional algorithms are replaced by models based on deep learning. 
Deep learning has proven to be successful in solving problems of large complexity, such as natural language processing or facial recognition tasks. 
Second, systems tend to be broken down into different components, in order to be placed where they are most needed and can be most efficient.
By establishing high-bandwidth connections between all kind of different devices and allowing many different system configurations, the components of the distributed system become part of what is known as the Internet of Things (IoT). \par
The aim of the VEDLIoT project is to provide tools, methodologies, and experience for supporting the development and deployment of distributed systems with components that rely on deep learning. 
The ability to decompose requirements and architecture appropriately is a key enabler for VEDLIoT. \par
Initially, the VEDLIoT toolchain and methods were applied to four use cases. One use case emerged from the automotive sector (automatic emergency braking), two from industry applications (fault detection in high voltage switches and operation condition classification for electric motors), and one use cases is in the field of smart homes (smart mirror). Over the course of the project, additional use cases will join the project as part of an open call.
\clearpage
\section{Mindmap created in the VEDLIoT workshop on challenges with distributed AI system}
\label{app:mindmap}
The mindmap depicted in Figure~\ref{fig:03_ClustersofConcern} was created during a joint industry - academic workshop on the challenges when designing distributed AI systems in VEDLIoT.
\begin{figure*}[!htbp]
    \centering
    \includegraphics[width=0.75\hsize]{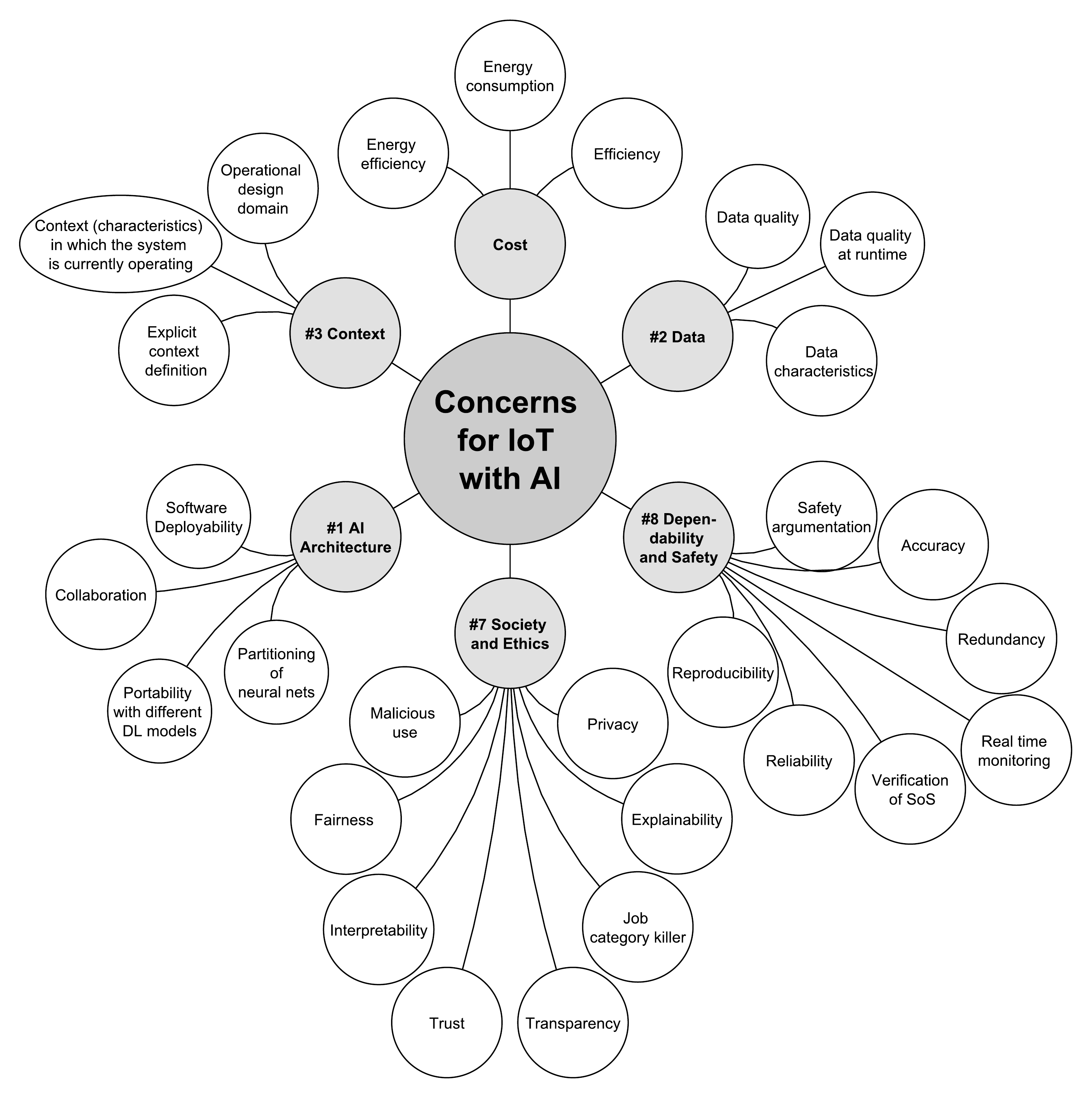}
    \caption{Concerns relevant for systems of the IoT with AI components.}
    \label{fig:03_ClustersofConcern}
\end{figure*}
\FloatBarrier

\section{Concepts from category theory compositional architecture thinking}
Category theory can be used to mathematically describe \emph{compositionality}, i.e., ``how systems or relationships can be combined to from new systems or relationships" \citep[Chapter 1]{Fong2018}. System and software architectures consist of architecture building blocks, which combined describe the desired systems \citep{Mitra2015}. \citet{Richards2020} suggests a "component-based thinking" for software architectures to easier communicate with developers and other stakeholders. Category theory can provide a mathematical foundation for a component-based construction of architectures, specifically supporting the co-design of different system aspects \citep{Zardini2020}. In our proposed compositional approach to an architecture framework, we utilise four concepts from category theory: \emph{Ordered sets}, \emph{Categories}, \emph{Morphisms}, and \emph{Functors}.

\paragraph*{Ordered sets}
\citet{Fong2018} state that ``category theory is all about organising and layering structures", which is why we apply the concept of ordered sets for differentiating and sorting architectural views into different levels of abstraction. A pre-order is a set of elements, that can be sorted such that a) \emph{reflexivity} and b)\emph{transitivity} holds, i.e., a) $a \leq a$ and b) if $a \leq b$ and $b \leq c$, then also $a \leq c$. It just says, that the elements in the set can be sorted by some logic, e.g., how detailed they describe a system aspect. Additionally, because a higher level of abstraction means a loss of details, i.e., information, the order is \emph{antisymmetric}\footnote{If you loose information from a more detailed system view, you cannot reconstruct it from the higher level of abstraction, but vice versa is possible}. Therefore, the set of architectural views in a cluster of concern are a \emph{partial order}.

\paragraph{Categories}
A category is a collection of elements, called objects. The type of object is irrelevant, but a pair of objects in a category is connected through \emph{morphisms}. For every object, there is also defined an \emph{identity morphism}. In our case, a category contains architectural views on the same level of abstraction.

\paragraph{Morphisms}
Morphisms exist between pairs of object (e.g. $a$ and $b$) in a category, denoted $f: a \rightarrow b$ and includes a set called "hom-set", indicated as $\text{Hom}(a;b)$. This "hom-set" could for example be a description how to "move" from one architectural view to another. For three (or more) objects $a,b,c$ in a category composition operations are allowed, i.e., $\text{Hom}(a;b) \times \text{Hom}(b;c) \rightarrow \text{Hom}(a;c)$.

\paragraph{Functors}
Given two categories $\mathbf{A}$ and $\mathbf{B}$, a functor $F$ maps all relations from $A$ to $B$. It allows us to keep correspondences defined on a higher level of abstraction intact when moving to the next lower level of abstraction.

\section{An antipattern in a compositional architecture framework}
The cybersecurity concept in Figure~\ref{fig:03_ArchViewpasorder} (on the conceptual level) should not have a direct (diagonal) correspondence rule to the component hardware architecture (on the design level). Instead, following Proposition~\ref{prop1}, the cybersecurity concept can have a relation to the technical cybersecurity concept (on the design level). According to Proposition~\ref{prop3}, this relation must then also exist on the next lower level of abstraction (i.e., between the technical cybersecurity concept and the component hardware architecture). If the higher level correspondence is not mapped onto lower levels, it is questionable why it exists at all (because the correspondence on the conceptual level has in that case no effect on the design level).
\clearpage

\section{Description of AI specific architectural views}
\label{app:descriptionarchiviews}
Table~\ref{05:tab_ArchiViews} contains descriptions and examples of architectural views that are specific for the AI component in the VEDLIoT project.
\begin{table*}[!htbp]
\centering
\caption{Description of AI specific architectural views. CoC: Cluster of Concern} 
\label{05:tab_ArchiViews}
\begin{singlespace}
\footnotesize
\makebox[\textwidth]{
\begin{tabular}{c|cll}
\hline
\rowcolor[HTML]{C0C0C0} 
\multicolumn{1}{c|}{\cellcolor[HTML]{C0C0C0}\textbf{CoC}} &
  \multicolumn{1}{c}{\textbf{View name}} &
  \multicolumn{1}{c}{\textbf{Description}} &
  \multicolumn{1}{c}{\textbf{Example}} \\ \hline

\parbox[t]{2mm}{\multirow{4}{*}[-4ex]{\rotatebox[origin=c]{90}{\textbf{Data Strategy}}}}
 &
  \begin{tabular}[c]{@{}c@{}}Data\\ ingestion\end{tabular} &
  \begin{tabular}[c]{@{}l@{}}Contains information on how  data is\\ going to be captured and stored.\end{tabular} &
  \begin{tabular}[c]{@{}l@{}}\rule{0pt}{2.5ex}Definition of a data storage structure\\ and necessary meta  information.\end{tabular} \\
 &
  \cellcolor[HTML]{EFEFEF}\begin{tabular}[c]{@{}c@{}}Data \\ selection \end{tabular} &
  \cellcolor[HTML]{EFEFEF}\begin{tabular}[c]{@{}l@{}}Provides rules on how the data is\\ selected for training. Defines ne-\\cessary data quality attributes.\end{tabular} &
  \cellcolor[HTML]{EFEFEF}\begin{tabular}[c]{@{}l@{}}\rule{0pt}{2.5ex}Specific features of the data are defined and used for filtering\\ the data set. Furthermore, data quality attributes are\\ set, such as image resolution and minimum brightness. \end{tabular} \\
\multicolumn{1}{l|}{} &
  \begin{tabular}[c]{@{}c@{}}Data preparation/ \\ manipulation\end{tabular} &
  \begin{tabular}[c]{@{}l@{}}Describes necessary preparations\\ and modifications on the dataset.\end{tabular} &
  \begin{tabular}[c]{@{}l@{}}\rule{0pt}{2.5ex}The camera can tilt, therefore the images are rotated.\\ Randomly, a percentage of the pixels are deactivated\\ to increase robustness of detection.\end{tabular} \\
  \multicolumn{1}{l|}{} &
  \cellcolor[HTML]{EFEFEF}\begin{tabular}[c]{@{}c@{}}\rule{0pt}{2.5ex}Run time data\\ monitoring \end{tabular} &
  \cellcolor[HTML]{EFEFEF}\begin{tabular}[c]{@{}l@{}}Concepts for ensuring that the data\\ specifications are met at run time.\end{tabular} &
  \cellcolor[HTML]{EFEFEF}\begin{tabular}[c]{@{}l@{}}A monitor is defined that checks the brightness level\\ of the image and the percentage of broken pixels.\end{tabular} \\
  
  \hline

\parbox[t]{2mm}{\multirow{4}{*}[-11ex]{\rotatebox[origin=c]{90}{\textbf{Learning}}}}
 &
  \begin{tabular}[c]{@{}c@{}}Learning\\ objectives\end{tabular} &
  \begin{tabular}[c]{@{}l@{}}High level architecture that out-\\lines which functions / models\\ shall be trained through learning.\end{tabular} &
  \begin{tabular}[c]{@{}l@{}}\rule{0pt}{2.5ex}One function is depicted that shall be able to recognise\\ pedestrians on the road. Three output classes are defined.\end{tabular} \\
 &
  \cellcolor[HTML]{EFEFEF}\begin{tabular}[c]{@{}c@{}}Learning\\ concept\end{tabular} &
  \cellcolor[HTML]{EFEFEF}\begin{tabular}[c]{@{}l@{}}Describes how the learning objective\\ shall be achieved. This can include\\ a description on how a simulation \\ environment is set up, which datasets\\ are required, if transfer learning\\ should be used, etc.\end{tabular} &
  \cellcolor[HTML]{EFEFEF}\begin{tabular}[c]{@{}l@{}}\rule{0pt}{2.5ex}The learning environment is a dataset containing\\ pedestrians on the road at different positions,\\ and empty roads. \end{tabular} \\
\multicolumn{1}{l|}{} &
  \begin{tabular}[c]{@{}c@{}}Learning\\ settings\end{tabular} &
  \begin{tabular}[c]{@{}l@{}}Detailed description of the learning\\ regime's configuration and setup.\end{tabular} &
  \begin{tabular}[c]{@{}l@{}}\rule{0pt}{2.5ex}Settings for learning, e.g. metric to use, stop condition,\\ loss function, optimiser, batch size, data split\\ for training and testing.\end{tabular} \\
  \multicolumn{1}{l|}{} &
  \cellcolor[HTML]{EFEFEF}\begin{tabular}[c]{@{}c@{}}\rule{0pt}{2.5ex}Manage continous \\improvements \end{tabular} &
  \cellcolor[HTML]{EFEFEF}\begin{tabular}[c]{@{}l@{}}A view on how continous im-\\provements, i.e., re-training,\\ can be achieved.\end{tabular} &
  \cellcolor[HTML]{EFEFEF}\begin{tabular}[c]{@{}l@{}}AI models can become outdated quickly through changing\\ environments. Continous data collection for\\ retraining can for example be specified here.\end{tabular} \\
  
  \hline
  
 \parbox[t]{2mm}{\multirow{4}{*}[-8ex]{\rotatebox[origin=c]{90}{\textbf{AI Model}}}}
 &
  \begin{tabular}[c]{@{}c@{}}High level\\ AI model\end{tabular} &
  \begin{tabular}[c]{@{}l@{}}\rule{0pt}{2.5ex}High Level architecture showing\\ which software components contain\\ an AI model. Also outlines, which \\ kind of AI is intended to be used.\end{tabular} &
  \begin{tabular}[c]{@{}l@{}}A deep neural network is assigned\\ to the function "Object detection".\end{tabular} \\
 &
  \cellcolor[HTML]{EFEFEF}\begin{tabular}[c]{@{}c@{}}AI model\\ concept\end{tabular} &
  \cellcolor[HTML]{EFEFEF}\begin{tabular}[c]{@{}l@{}}Setup of the AI model\\ outlining the architecture of the AI.\end{tabular} &
  \cellcolor[HTML]{EFEFEF}\begin{tabular}[c]{@{}l@{}}\rule{0pt}{2.5ex}For object detection using a sequential deep neural\\ network with a number of 2D-convolution layers,\\ a 2D-pooling layer, dropout, flatten, and a\\ dense layer. Input, and output definition.\end{tabular} \\
\multicolumn{1}{l|}{} &
  \begin{tabular}[c]{@{}c@{}}AI model\\ configuration\end{tabular} &
  \begin{tabular}[c]{@{}l@{}}Detailed description of the \\ AI model configuration.\end{tabular} &
  \begin{tabular}[c]{@{}l@{}}\rule{0pt}{2.5ex}For a deep neural network this can be kernel size of\\ convolution layers, dropout rate, activation function, etc.\end{tabular} \\ 
\multicolumn{1}{l|}{} &
  \cellcolor[HTML]{EFEFEF}\begin{tabular}[c]{@{}c@{}}\rule{0pt}{2.5ex}AI model \\ performance \\ monitoring \end{tabular} &
  \cellcolor[HTML]{EFEFEF}\begin{tabular}[c]{@{}l@{}}Concept for monitoring the \\ performance of the AI.\end{tabular} &
  \cellcolor[HTML]{EFEFEF}\begin{tabular}[c]{@{}l@{}}For example recording of inference time \\ and uncertainty monitoring.\end{tabular} \\

  \hline
 
 \parbox[t]{2mm}{\multirow{4}{*}[0ex]{\rotatebox[origin=c]{90}{\textbf{Context \& Constraints}}}}
 &
  \begin{tabular}[c]{@{}c@{}}\rule{0pt}{2.5ex}Context\\ assumptions\end{tabular} &
  \begin{tabular}[c]{@{}l@{}}Assumption on the context \\ (this can be the environment, \\ involved users, etc.).\end{tabular} &
  \begin{tabular}[c]{@{}l@{}}Object detection shall operate \\ at all times on Swedish roads.\end{tabular} \\
 &
  \cellcolor[HTML]{EFEFEF}\begin{tabular}[c]{@{}c@{}}Context\\ definition\end{tabular} &
  \cellcolor[HTML]{EFEFEF}\begin{tabular}[c]{@{}l@{}}Detailed description of the \\ context in which the desired \\ behaviour will be executed.\end{tabular} &
  \cellcolor[HTML]{EFEFEF}\begin{tabular}[c]{@{}l@{}}\rule{0pt}{2.5ex}Swedish highways and motorways. All possible weather\\ conditions in Sweden. Behaviour should be stable\\ at all times of the day and night.\end{tabular} \\
\multicolumn{1}{l|}{} &
  \begin{tabular}[c]{@{}c@{}}Constraints /\\ Design\\ Domain\end{tabular} &
  \begin{tabular}[c]{@{}l@{}}Lists constraints, and design domain,\\ in which the desired behaviour will\\ be expected.\end{tabular} &
  \begin{tabular}[c]{@{}l@{}}\rule{0pt}{2.5ex} Object to detect is not more than 1 metre above the road.\\ At night, headlights illuminate  road at least 20 metres ahead.\end{tabular} \\
\multicolumn{1}{l|}{} &
  \cellcolor[HTML]{EFEFEF}\begin{tabular}[c]{@{}c@{}}Context \\ monitoring \end{tabular} &
  \cellcolor[HTML]{EFEFEF}\begin{tabular}[c]{@{}l@{}}Concept for monitoring \\ that the AI operates in the \\ desired context of operation.\end{tabular} &
  \cellcolor[HTML]{EFEFEF}\begin{tabular}[c]{@{}l@{}}\rule{0pt}{2.5ex}System monitors the geographical location of the vehicle\\ (e.g., through GPS) and prevents activation of system if outside\\ of desired context (e.g., Swedish highways and motorways).\end{tabular} \\  
  
  \hline
\end{tabular}}
\end{singlespace}
\end{table*}
\end{onecolumn}

\clearpage
\section{Table of morphisms / relations between architectural views} \label{app:relations}
Table~\ref{tab:corr_example} provides examples of correspondences between architectural views in the VEDLIoT architecture framework. The examples are from an extract of the framework used for the automotive use case illustrated in Figure~\ref{fig:04_UC2_CLAI}. The correspondences are visualised in Figure~\ref{fig:04_UC2_CLAI_corr}.

\begin{table}[h!]

\centering
\caption{Morphisms / relations between the architecture views in the VEDLIoT use case architecture framework extract illustrated Figure~\ref{fig:04_UC2_CLAI}. LoA: Level of Abstraction, A: Analytical Level, B: Conceptual Level, C: Design Level, D: Run time Level}
\label{tab:corr_example}
\tiny
\begin{tabular}{llllllll}
\hline
\rowcolor[HTML]{C0C0C0} 
\textbf{ID}   & \textbf{LoA} & \textbf{Origin view}      & \textbf{Effecting component}       & \textbf{Target view}                                                                & \textbf{Affected component}        & \textbf{Description}                                                                              & \textbf{Inherited} \\ \hline
A-01 & A   & Context assumptions                                                        & context description                                                  & Training objectives                                                        & classification categories & Definition of pedestrians                                                                &           \\ \rowcolor[HTML]{EFEFEF}
A-02 & A   & Training objectives                                                        & data labels                                                          & Data ingestion                                                             & strings of data labels    & Names of labels                                                                          &           \\
A-03 & A   & Context assumptions                                                        & context description                                                  & High level AI model                                                        & output definition         & Number of output classes                                                                 &           \\ \rowcolor[HTML]{EFEFEF}
B-01 & B   & Context definition                                                         & context description                                                  & Training concept                                                           & classification categories & \begin{tabular}[c]{@{}l@{}}Definition of \\ pedestrians\end{tabular}                     & A-01      \\
B-02 & B   & Training concept                                                           & data labels                                                          & Data selection                                                             & strings of data labels    & Names of labels                                                                          & A-02      \\ \rowcolor[HTML]{EFEFEF}

B-03 & B   & Context definition                                                        & context description                                                  & AI model concept                                                           & output layer              & \begin{tabular}[c]{@{}l@{}}number of neurons,\\ type of output classes\end{tabular}      & A-03      \\
B-04 & B   & \begin{tabular}[c]{@{}l@{}}System hardware \\ architecture\end{tabular}                                                        & camera                                                  & Context definition                                                           & context description              & \begin{tabular}[c]{@{}l@{}}Position of camera \\ min. height of pedestrian\end{tabular}      &      \\\rowcolor[HTML]{EFEFEF}

B-05 & B   & \begin{tabular}[c]{@{}l@{}}System hardware \\ architecture\end{tabular}    & camera                                                               & Data selection                                                             & data quality attributes   & \begin{tabular}[c]{@{}l@{}}input data type, \\ resolution\end{tabular}                   &           \\ 
B-06 & B   & Context definition                                                         & context description                                                  & Data selection                                                             & feature attributes        & feature descriptions                                                                     &           \\ \rowcolor[HTML]{EFEFEF}
B-07 & B   & \begin{tabular}[c]{@{}l@{}}System hardware \\ architecture\end{tabular}    & camera                                                               & AI model concept                                                           & input layer               & \begin{tabular}[c]{@{}l@{}}input data type, input \\ layer size, resolution\end{tabular} &           \\ 
B-08 & B   & AI model concept                                                           & size of DNN                                                          & \begin{tabular}[c]{@{}l@{}}System hardware\\  architecture\end{tabular}    & processing unit           & \begin{tabular}[c]{@{}l@{}}memory and processing \\ needs\end{tabular}                   &           \\ \rowcolor[HTML]{EFEFEF}
C-01 & C   & Design domain                                                              & context description                                                  & Optimiser settings                                                         & classification categories & \begin{tabular}[c]{@{}l@{}}Definition of \\ pedestrians\end{tabular}                      & B-01      \\ 
C-02 & C   & Optimiser settings                                                         & data labels                                                          & Data preparation                                                           & strings of data labels    & Names of labels                                                                          & B-02      \\\rowcolor[HTML]{EFEFEF}
C-03 & C   & Design domain                                                              & context description                                                  & \begin{tabular}[c]{@{}l@{}}AI model \\ configuration\end{tabular}          & output layer              & \begin{tabular}[c]{@{}l@{}}number of neurons, \\ type of output classes\end{tabular}     & B-03      \\ 
C-04 & C   & \begin{tabular}[c]{@{}l@{}}Component hardware \\ architecture\end{tabular}                                                        & camera                                                  & Design domain                                                           & context description              & \begin{tabular}[c]{@{}l@{}}Position of camera \\ min. height of pedestrian\end{tabular}      &  B-04    \\\rowcolor[HTML]{EFEFEF}

C-05 & C   & \begin{tabular}[c]{@{}l@{}}Component hardware \\ architecture\end{tabular} & camera                                                               & Data preparation                                                           & data quality attributes   & \begin{tabular}[c]{@{}l@{}}input data type, \\ resolution\end{tabular}                   & B-05      \\
C-06 & C   & Design domain                                                              & context description                                                  & Data preparation                                                           & feature attributes        & feature descriptions                                                                     & B-06      \\ \rowcolor[HTML]{EFEFEF}
C-07 & C   & \begin{tabular}[c]{@{}l@{}}Component hardware \\ architecture\end{tabular} & camera                                                               & \begin{tabular}[c]{@{}l@{}}AI model \\ configuration\end{tabular}          & input layer               & \begin{tabular}[c]{@{}l@{}}input data type, \\ resolution\end{tabular}                   & B-07      \\
C-08 & C   & \begin{tabular}[c]{@{}l@{}}AI model\\ configuration\end{tabular}           & size of DNN                                                          & \begin{tabular}[c]{@{}l@{}}Component hardware \\ architecture\end{tabular} & Jetson NX                 & \begin{tabular}[c]{@{}l@{}}memory and processing \\ needs\end{tabular}                   & B-08      \\ \rowcolor[HTML]{EFEFEF}
C-09 & C   & \begin{tabular}[c]{@{}l@{}}Component hardware \\ architecture\end{tabular} & Jetson NX                                                            & \begin{tabular}[c]{@{}l@{}}AI model \\ configuration\end{tabular}          & configuration of DNN      & \begin{tabular}[c]{@{}l@{}}model optimisation \\ for hardware\end{tabular}               &           \\
C-10 & C   & \begin{tabular}[c]{@{}l@{}}Component hardware \\ architecture\end{tabular} & camera                                                               & Data preparation                                                           & data manipulation         & error rate of pixels                                                                     &           \\ \rowcolor[HTML]{EFEFEF}
C-11 & C   & Design domain                                                              & \begin{tabular}[c]{@{}l@{}}design domain \\ description\end{tabular} & Data preparation                                                           & data manipulation         & \begin{tabular}[c]{@{}l@{}}random tilting \\ of input image\end{tabular}                 &           \\
C-12 & C   & \begin{tabular}[c]{@{}l@{}}AI model \\ configuration\end{tabular}          & type of DNN                                                          & Optimiser settings                                                         & Optimiser setings         & \begin{tabular}[c]{@{}l@{}}set of parameters \\ for optimiser\end{tabular}               &           \\ \rowcolor[HTML]{EFEFEF}
D-01 & D   & Context monitoring                                                         & context description                                                  & \begin{tabular}[c]{@{}l@{}}Continous \\ improvements\end{tabular}          & classification categories & \begin{tabular}[c]{@{}l@{}}Definition of \\ pedestrians\end{tabular}                     & C-01      \\
D-02 & D   & \begin{tabular}[c]{@{}l@{}}Continous \\ improvements\end{tabular}          & data labels                                                          & \begin{tabular}[c]{@{}l@{}}Run time data \\ monitoring\end{tabular}         & strings of data labels    & Names of labels                                                                          & C-02      \\ \rowcolor[HTML]{EFEFEF}
D-03 & D   & Context monitoring                                                         & context description                                                  & \begin{tabular}[c]{@{}l@{}}AI performance \\ monitoring\end{tabular}       & uncertainty monitoring    & \begin{tabular}[c]{@{}l@{}}number of neurons, \\ type of output classes\end{tabular}     & C-03      \\
D-04 & D   & \begin{tabular}[c]{@{}l@{}}Hardware performance \\ monitoring\end{tabular}                                                        & camera                                                  & Context monitoring                                                           & context description              &  \begin{tabular}[c]{@{}l@{}}Position of camera \\ min. height of pedestrian\end{tabular}      &  C-04 \\ \rowcolor[HTML]{EFEFEF}
D-05 & D   & \begin{tabular}[c]{@{}l@{}}Hardware performance \\ monitoring\end{tabular} & camera                                                               & \begin{tabular}[c]{@{}l@{}}Run time data \\ monitoring\end{tabular}         & check resolution          & \begin{tabular}[c]{@{}l@{}}input data type, \\ resolution\end{tabular}                   & C-05      \\ 
D-06 & D   & Context monitoring                                                         & context description                                                  & \begin{tabular}[c]{@{}l@{}}Run time data \\ monitoring\end{tabular}         & data testing              & feature descriptions                                                                     & C-06      \\ \rowcolor[HTML]{EFEFEF}
D-07 & D   & \begin{tabular}[c]{@{}l@{}}Hardware performance \\ monitoring\end{tabular} & camera                                                               & \begin{tabular}[c]{@{}l@{}}AI performance \\ monitoring\end{tabular}       & test and validation       & \begin{tabular}[c]{@{}l@{}}input data type, \\ resolution\end{tabular}                   & C-07      \\ 
D-08 & D   & \begin{tabular}[c]{@{}l@{}}AI performance \\ monitoring\end{tabular}       & size of DNN                                                          & \begin{tabular}[c]{@{}l@{}}Hardware performance \\ monitoring\end{tabular} & hardware watchdog         & \begin{tabular}[c]{@{}l@{}}memory and processing \\ needs\end{tabular}                   & C-08      \\ \rowcolor[HTML]{EFEFEF}
D-09 & D   & \begin{tabular}[c]{@{}l@{}}Hardware performance \\ monitoring\end{tabular} & Jetson NX                                                            & \begin{tabular}[c]{@{}l@{}}AI performance \\ monitoring\end{tabular}       & test and validation       & \begin{tabular}[c]{@{}l@{}}model optimisation \\ for hardware\end{tabular}               & C-09      \\ 
D-10 & D   & \begin{tabular}[c]{@{}l@{}}Hardware performance \\ monitoring\end{tabular} & camera                                                               & \begin{tabular}[c]{@{}l@{}}Run time data \\ monitoring\end{tabular}         & check broken pixels       & error rate of pixels                                                                     & C-10      \\ \rowcolor[HTML]{EFEFEF}
D-11 & D   & Context monitoring                                                         & \begin{tabular}[c]{@{}l@{}}design domain \\ description\end{tabular} & \begin{tabular}[c]{@{}l@{}}Run time data \\ monitoring\end{tabular}         & test and validation       & \begin{tabular}[c]{@{}l@{}}random tilting of \\ input image\end{tabular}                 & C-11      \\ 
D-12 & D   & \begin{tabular}[c]{@{}l@{}}AI performance \\ monitoring\end{tabular}       & type of DNN                                                          & \begin{tabular}[c]{@{}l@{}}Manage continous \\ improvements\end{tabular}   & Optimiser setings         & \begin{tabular}[c]{@{}l@{}}set of parameters \\ for optimiser\end{tabular}               & C-12      \\ \rowcolor[HTML]{EFEFEF}
D-13 & D   & \begin{tabular}[c]{@{}l@{}}Run time data \\ montoring\end{tabular}          & check input image                                                    & \begin{tabular}[c]{@{}l@{}}Continous \\ improvements\end{tabular}          & capturing of new data     & \begin{tabular}[c]{@{}l@{}}quality requirements \\ towards new images\end{tabular}       &          
\end{tabular}
\end{table}





\end{document}